\documentclass[10pt,journal,twoside,twocolumn]{IEEEtran}

\def\psfancypar#1#2{\begingroup\def\par{\endgraf\endgroup\lineskiplimit=0pt}
               \setbox2=\hbox{\large\sc #2}
               \newdimen\tmpht \tmpht \ht2 \advance\tmpht by \baselineskip
               \font\hhuge=Times-Bold at \tmpht
               \setbox1=\hbox{{\hhuge #1}}
               \count7=\tmpht \count8=\ht1
               \divide\count8 by 1000 \divide\count7 by \count8 
               \tmpht=.001\tmpht\multiply\tmpht by \count7 
               \font\hhuge=Times-Bold at \tmpht
               \setbox1=\hbox{{\hhuge #1}}
               \noindent
                \hangindent1.05\wd1
               \hangafter=-2 {\hskip-\hangindent
               \lower1\ht1\hbox{\raise1.0\ht2\copy1}%
                \kern-0\wd1}\copy2\lineskiplimit=-1000pt}

\newcommand{\Gammabf}{\mbox{${\bf \Gamma}$}}

\newcommand{\E}{\mbox{{\rm E}}}

\newcommand{\abf}{\mbox{${\bf a}$}}

\def\boxit#1{\vbox{\hrule\hbox{\vrule\kern3pt
        \vbox{\kern3pt#1\kern3pt}\kern3pt\vrule}\hrule}}

\def\reals{ { {\rm  I \kern-0.15em R }  } }
\def\complex{ {\,{{\rm C} \kern-0.50em \raise0.20ex {  |}}\, }}

\def\lambdabf{\hbox{\boldmath$\lambda$\unboldmath}}

\def\xibf{\hbox{\boldmath$\xi$\unboldmath}}

\def\rhobf{\hbox{\boldmath$\rho$\unboldmath}}

\def\Sigmabf{\hbox{$\bf \Sigma$}}

\def\Gammabf{\hbox{$\bf \Gamma$}}

\def\Lambdabf{\mbox{$ \bf \Lambda $}}

\def\abf{{\bf a}}
\def\bbf{{\bf b}}

\def\ebf{{\bf e}}

\def\hbf{{\bf h}}

\def\pbf{{\bf p}}

\def\sbf{{\bf s}}

\def\ubf{{\bf u}}

\def\wbf{{\bf w}}

\def\ybf{{\bf y}}

\def\ybf{{\bf y}}
\def\Abf{{\bf A}}
\def\Bbf{{\bf B}}
\def\Cbf{{\bf C}}
\def\Dbf{{\bf D}}

\def\Fbf{{\bf F}}

\def\Hbf{{\bf H}}
\def\Ibf{{\bf I}}

\def\Kbf{{\bf K}}

\def\Pbf{{\bf P}}

\def\Rbf{{\bf R}}
\def\Sbf{{\bf S}}

\def\Ubf{{\bf U}}
\def\Vbf{{\bf V}}

\def\Hc{{\cal H}}
\def\Ic{{\cal I}}

\def\Kc{{\cal K}}
\def\Lc{{\cal L}}

\def\Sc{{\cal S}}

\def\be{\vskip .3cm \begin{equation}}
\def\ee{\end{equation} \vskip .4cm \noindent}
\newcommand{\R}{\mbox{$\hat {\bf R}_{N}$}}

\def\Rxx{\Rbf_{\ssstyle X\kern-.1em X}}

\let\ssstyle=\scriptscriptstyle

\def\Kout{\setbox1=\hbox{\Huge\bf K}\hbox to
1.05\wd1{\hspace{.05\wd1}
}}
\def\Sout{\setbox1=\hbox{\Huge\bf S}\hbox to 1.05\wd1{\hspace{.05\wd1}
}}

  \ifx\LabelFigloaded\MYundefined\relax
  \else
   \fi

  \def\LabelFigloaded{\relax}

  \chardef\LabelFigCatAt\the\catcode`\@
  \catcode`\@=11

 \let\LabelFigwlog@ld\wlog
 \def\wlog#1{\relax}

 \ifx\\\MYundefined@
    \let\\\relax
 \fi

  \def\ms@g{\immediate\write16}

 \def\N@wif{\csname newif\endcsname }
 \def\Temp@ {\N@wif\ifIN@}
 \ifx\INN@\MYundefined@
    \else \let\Temp@\relax
 \fi
 \Temp@

  \def\IN@{\expandafter\INN@\expandafter}
  \long\def\INN@0#1@#2@{\long\def\NI@##1#1##2##3\ENDNI@
    {\ifx\m@rker##2\IN@false\else\IN@true\fi}%
     \expandafter\NI@#2@@#1\m@rker\ENDNI@}
  \def\m@rker{\m@@rker}
 
  \newtoks\Initialtoks@  \newtoks\Terminaltoks@
  \def\SPLIT@{\expandafter\SPLITT@\expandafter}
  \def\SPLITT@0#1@#2@{\def\TTILPS@##1#1##2@{%
     \Initialtoks@{##1}\Terminaltoks@{##2}}\expandafter\TTILPS@#2@}

 \def\Shifted@@#1#2#3{\setbox0=\hbox{#3}%
   \raise -\dp0\vbox {\kern-#2%
       \hbox {\kern#1\unhbox0\kern-#1}%
           \kern#2}}

 \newcount\gridcount
 \newbox\auxGridbox@ \newbox\hGridbox@ \newbox\vGridbox@
 \newbox\Labelbox@ \newbox\auxLabelbox@
 \newbox\Coordinatebox@
 \newtoks\Labeltoks@
 \newdimen\Wdd@ \newdimen\Htt@
 \newdimen\Wddd@ \newdimen\Httt@
 
 \def\Wr@{\immediate\write16}

 \newdimen\GL@wd
 \def\GridLineWidth#1{\GL@wd=#1}

 \def\gobble#1{}
 \def\EdgeErr@{\Wr@{}%
      \Wr@{\string\Edges\space argument
      1, 10, 100 or 1000 please\string!}%
      }

 \newcount\Edgect@

 \def\Sweepup#1\endSweepup{}

 \def\SetEdges@{%
    \edef\Zr@@s{\expandafter\gobble\number\Edgect@\empty}%
        \count255=0\Zr@@s\relax
        \ifnum\count255=\z@\else\EdgeErr@\show\tailtest\fi
        \count255=1\Zr@@s\relax
        \ifnum\count255=\Edgect@\relax\else\EdgeErr@\show\leadtest\fi
    \EdgGl@b\edef\Zr@s{\expandafter\gobble\Zr@@s\empty}
    \ifnum\Edgect@>\@ne\relax\EdgGl@b\let\L@Dc\empty
        \else\EdgGl@b\edef\L@Dc{\string.}\fi
    \ifnum\Edgect@>\@ne\relax
        \EdgGl@b\edef\Edgescale@##1{\divide##1 by \Edgect@}%
        \else\EdgGl@b\edef\Edgescale@##1{}\fi
    }

 \def\Edges#1{\Edgect@=#1\relax
     \let\EdgGl@b\global \SetEdges@}

 \Edges{1}

 \def\hhrule{\hrule height \GL@wd\vskip-.\GL@wd}

 \def\hRule@{%
   \advance\gridcount -2%
   \vfil\hhrule\vfil
   \llap{\smash{\raise -2.5pt
     \hbox{\L@Dc\number\gridcount\Zr@s\kern2pt}}}%
   \hhrule
   }

\def\vvrule{\vrule width \GL@wd \kern-\GL@wd}

 \def\vRule@{\advance\gridcount 2%
   \hfil\vvrule\hfil
   \setbox\auxGridbox@=\vbox to 0pt
      {\vskip \Htt@\vskip 2pt
        \hbox to 0pt{\hss\L@Dc\number\gridcount\Zr@s\hss}\vss}%
      \wd\auxGridbox@=0pt \box\auxGridbox@
   \vvrule
   }

 \def\PlaceGrid@@{\gridcount=10 
  \setbox\hGridbox@=\hbox{%
        \hbox{%
             \hskip-.4pt\vrule
             \vbox to \Htt@{%
               \offinterlineskip\parindent=\z@\relax
               \hbox to \Wdd@{\hfil}
               \hRule@\hRule@\hRule@\hRule@
               \vfil\hhrule\vfil}%
             \vrule\hskip-.4pt}
    }%
  \gridcount=0%
  \setbox\vGridbox@=\hbox{%
      \vbox{\offinterlineskip\parindent=0pt\hsize=0pt
         \vskip-.4pt\hrule%
         \hbox to \Wdd@{%
                 \vtop to \Htt@{\vfil}%
                 \vRule@\vRule@\vRule@\vRule@
                 \hfil\vvrule\hfil}%
         \hrule\vskip-.4pt}}%
  \wd\hGridbox@=0pt\ht\hGridbox@=0pt
  \wd\vGridbox@=0pt\ht\vGridbox@=0pt
  \hbox{\box\hGridbox@\box\vGridbox@}%
  }

 \def\LabelsGlobal{\def\LabGl@b{\global}}
 \def\LabelsLocal{\def\LabGl@b{}}
 \LabelsGlobal 

 \def\SetLabels#1\endSetLabels{%
   \LabGl@b\Labeltoks@={#1()\\}%
   }

 \LabGl@b\Labeltoks@={()\\}

 \def\ShowGrid{\LabGl@b\let\PlaceGrid@\PlaceGrid@@}
 \def\HideGrid{\LabGl@b\let\PlaceGrid@\relax}
 \def\Grids{\ShowGrid\LabGl@b\let\GridSwitch@\ShowGrid}
 \def\noGrids{\HideGrid\LabGl@b\let\GridSwitch@\HideGrid}

 \noGrids

 \def\bAdjust@@{%
     \setbox\auxLabelbox@=\hbox{\raise \dp\auxLabelbox@
            \box\auxLabelbox@}}
 \def\bAdjust@{\let\vAdjust@\bAdjust@@}

 \def\eAdjust@@{\dimen0=-.5\ht\auxLabelbox@
     \advance\dimen0 by .5\dp\auxLabelbox@
     \setbox\auxLabelbox@=
            \hbox{\raise\dimen0\box\auxLabelbox@}}
 \def\eAdjust@{\let\vAdjust@\eAdjust@@}

 \def\tAdjust@@{%
     \setbox\auxLabelbox@=\hbox{\raise-\ht\auxLabelbox@
            \box\auxLabelbox@}}
 \def\tAdjust@{\let\vAdjust@\tAdjust@@}

 \let\vAdjust@\relax

 \def\lAdjust@{\let\hAdjust@\rlap}
 \def\rAdjust@{\let\hAdjust@\llap}

 \let\hAdjust@\relax\let\vAdjust@\relax

 \def\FetchLabel@#1(#2)#3\\{%
     \IN@0#2@@\ifIN@
        \setbox0=\hbox{\ignorespaces#1#3\unskip}%
        \ifdim\wd0>0pt
           \ms@g{}%
           \ms@g{ !!! Bad label(s)? !!!}%
           \message{ #1(#2)#3}%
        \fi
        \def\LabelMole@##1\endFetchLabel@{%
            \IN@0()\\@##1@%
            \ifIN@\def\Temp@{\FetchLabel@##1\endFetchLabel@}%
            \else\def\Temp@{}%
            \fi
            \Temp@
           }%
     \else
       \ignorespaces#1\unskip
       \setbox\auxLabelbox@=%
         \hbox to 0pt{\hss\ignorespaces\hAdjust@
          {\ignorespaces#3\unskip}\hss}%
       \vAdjust@
       \let\hAdjust@\relax\let\vAdjust@\relax
       \AugmentLabelBox@@{#2}%
       \ht\Labelbox@=0pt\dp\Labelbox@=0pt
       \let\LabelMole@\FetchLabel@%
     \fi\LabelMole@}

 \newtoks\XYSep@ 
 \def\SetXYSeparator#1{%
     \IN@0#1@@\ifIN@\XYSep@{*}%
     \else
     \XYSep@{#1}%
     \fi
     }

 \SetXYSeparator*

 \def\AugmentLabelBox@@#1{%
     \IN@0\the\XYSep@ @#1@\ifIN@
       \SPLIT@0\the\XYSep@ @#1@%
       \setbox\Labelbox@=\hbox to 0pt{%
         \unhbox\Labelbox@
         \Shifted@@{\the\Initialtoks@\Wddd@}%
         {\the\Terminaltoks@\Httt@}%
         {\box\auxLabelbox@}}%
     \else
         \ms@g{}%
         \ms@g{ !!! Bad insertion point. !!!}%
         \message{ (#1\ this point was rejected.)}%
     \fi
    }

 \def\FetchOption@#1[#2]#3\endFetchOption@{%
    \def\temp{#1}
    \ifx\temp\empty
       \Edgect@=#2\relax
       \let\EdgGl@b\relax
       \SetEdges@
       \Cleaner@#3%
    \fi}

 \def\Cleaner@#1[@]{\Labeltoks@{#1}}
     
 \def\PlaceLabels@@{\mathsurround=0pt
     \def\Cr@{\\}%
     \let\L\lAdjust@\let\R\rAdjust@
     \let\B\bAdjust@\let\E\eAdjust@\let\T\tAdjust@
     \expandafter\FetchOption@\the\Labeltoks@[@]\endFetchOption@
     \Wddd@=\Wdd@ \Edgescale@\Wddd@ 
     \Httt@=\Htt@ \Edgescale@\Httt@
     \expandafter\FetchLabel@\the\Labeltoks@\endFetchLabel@
     \box\Labelbox@
     }%

 \let \PlaceLabels@\PlaceLabels@@

 \def\AffixLabels#1{\setbox\Coordinatebox@=\hbox{#1}%
      \Wdd@=\wd\Coordinatebox@ \Htt@=\ht\Coordinatebox@
      \advance\Htt@ \dp\Coordinatebox@
      \hbox{\copy\Coordinatebox@\kern-\Wdd@ 
           \Shifted@@{0pt}{-\dp\Coordinatebox@}%
           {\PlaceLabels@\PlaceGrid@}%
           \kern\Wdd@}%
      \GridSwitch@ 
      \LabGl@b\Labeltoks@{()\\}%
      }
 
   \let\wlog\LabelFigwlog@ld   
   \catcode`\@=\LabelFigCatAt  

                                By

              Raymond S\'eroul <A18645@FRCCSC21.BITNET>
                                and 
              Laurent Siebenmann <lcs@topo.math.u-psud.fr>
    
              VERSIONS: July 1991, Oct 1991, Jan 1992, July 1992

INTRODUCTION

      This labelling package is intended for TeX users who
rely on non-TeX sources for for their graphics inserts.  It
provides means for adding TeX labels to such inserts with a
minimum of fuss. 

       For most labels, TeX users have in the past found it
reasonably convenient to rely on non-TeX sources. Typical
occasions when an inescapable need for TeX labels seemed to
arise are

 (a) when the graphics program lacks certain exotic or complex
mathematical symbols

 (b) when the very highest typographical quality is wanted for the
labels

 (c) when labels included with the graphics fail to print, 
 and you cannot figure out why (cf. boxedeps.doc).  The labels
 provided by labelfig.tex are 100

       Since this package first appeared, many users, who in the
past scarcely dreamed of using TeX labels, have come to use
nothing but.  So it is now appropriate to add

Intoxication Warning:  TeX labels may be addictive and expensive. 

     If you have a fast preview you may disagree, and even find
that this package provides an agreeable paste-up environment; see
extra applications at end.

     Note to publishers: It is possible and convenient to ultimately
export the TeX labels produced by labelfig.tex to become an integral
part of the EPS file. This is often desired by a publisher who typically
uses an "upmarket" graphics or page layout program, with which the
staff is skilled in perfecting figures.  See Appendix I for
a recipe.

     The authors are grateful to Patrick Ion of Math Reviews for
helpful comments and encouragement.

BASIC INSTRUCTIONS

    After reading in the macro file using

preview or proof your figure with a coordinate grid printed on
top, by typing the following:

    \ShowGrid  
    \AffixLabels{<the graphics insertion>}

Here <the graphics insertion> is what you would type to insert
the graphics object alone without the grid.  This must provide
for the space around it. For example <the graphics insertion>
might well be \BoxedEPSF{MyFigure scaled 700} using the
boxedeps.tex macro package (from same source); this provides a
TeX box containing the encapsulated PostScript insert specified by
the file MyFigure. \AffixLabels{...} provides the grid (supposing
\ShowGrid is present) and later, once you have specified labels
using the grid, it will "tack on" the labels.

     The grid is a sort of (usually elongated) checkerboard of
ten rows and ten columns and its (internal) partitions are by
default numbered  .1, ... ,.9  both horizontally (X-coordinate
running left to right) and vertically (Y-coordinate running bottom
to top).  Thus the points enclosed by the grid correspond to the
points of the unit square in the cartesian "X-Y" plane, the lower
left corner corresponding to the origin (0,0).  By extrapolation,
the full page corresponds to a larger rectangle in the plane.

     These coordinates serve to position labels as follows.
Before the \AffixLabels{...} command type label specifications:

  \SetLabels
   (<X-coordinate>*<Y-coordinate>) <first label> \\
   .
   .
   .
   (<X-coordinate>*<Y-coordinate>)  <last label> \\
  \endSetLabels

Each row specifies one label and is terminated by \\.  In each
row, the position indicator comes first; it is written as a
standard cartesian point except that the X- and Y- coordinates
are separated by * rather than a comma because TeX allows a
comma as decimal point. There are no dimension units to specify
as the unit is the grid itself.

     By default, this cartesian point specifies where the middle
of the baseline of the label will be located.  However if you precede
the point by \L [or \R] the left [or right] edge of the baseline will
be located there. Similarly you may also precede the point by \T, \E,
or \B to vertically align the top equator or bottom of the label box
at the specified point.  This gives nine standard positions of
the label with respect to the insertion point --- corresponding to
the eight principle points of the compas and the center

                     \L\T     \T      \R\T

                     \L\E     \E      \R\E

                     \L\B     \B      \R\B

But this neglects the default "baseline" level of TeX,
giving potentially three more positions

                     \L    <no tag>   \R

For text, the baseline level is often the preferred. Its relation to
the others is variable. It will often coincide with the bottom level,
as happens for "X".  But it is often distinct, as for "g", in which
case you have in all 12 distinct positions rather than 9.

     It is convenient to think of this specification of label
position as attaching the label by a thumb-tack to the coordinate
grid. There are up to twelve positions of the thumb-tack on the
label, while the position of the thumb-tack on the coordinate grid is
arbitrary.  Normally, one choses the position of the thumb-tack on
the label to be the one that is the closest to the item being
labeled.  There are good reasons for this "rule of thumb":

   (a)  It facilitates correct positioning at first try.

   (b)  If the scale of the figure must be altered after labels
have been affixed, the labels have a good chance of remaining well
positioned.

   (c)  The visible grid need not extend beyond the "bounding box"
for the figure, because the best preferred position is always
(at least almost) within the bounding box .

The second reason is particularly important. Indeed it often
happens that scale has to be altered after labelling begins, in
order to either provide space for the labels, or to adjust
proportions between the labels and the figure.  (The size of labels
is unaffected by scaling.)

     Here is an artificial but self-contained test which uses
TeX rules to make a graphics object.

TEST

    Do not skip this!

 \def\FrameIt#1{\hbox{\vrule$\vcenter {\hrule\kern3pt%
             \hbox {\kern3pt #1\kern3pt}%
               \kern3pt\hrule}$\relax\vrule}}

 \def\Caption#1#2{\FrameIt{%
       \vtop {\hsize=#1\relax \parindent=0pt
         \leftskip=0pt \rightskip=0pt plus15pt
         \parfillskip=0pt
         \lineskip=1pt\baselineskip=0pt
         #2}}}

 \def\FirstQuadrant{\hbox to 100pt{\vrule\vbox to 100pt{%
        \hbox to 100pt{\hfil}\vfil\hrule}\hss}}

  \SetLabels
    \R(.5*.2) $\zeta\,\cdot$\\
    (.9*-.10) $\xi$\\
    \R(-.03*.9) $\eta$\\
    \T(.5*.9) \Caption{70pt}{%
          \it The norm of
          $g(\xi+i\eta)$ is indicated on
          contours of this invisible surface.}\\
  \endSetLabels

  \AffixLabels{\FirstQuadrant}

  \end

  Note that the coordinates to use for labels are indicated on the
edges of the grid (when visible) corresponding to the conventional
x- and y- axes of the Cartesian plane. By default the grid is
1-by-1. However, by the command \Edges{100}, you can change this
to 100-by-100 and many users find this alternative most
convenient. Place the command \Edges{...} in your style file (or
header) since its effect is is global. Other possible edge values
are 10 and 1000.

  If you use the command \Edges{...} at all, do so with care.  For
if you accidentally delete an \Edges{...} command your labels will
abruptly be badly misplaced and may logically but mysteriously
generate "dimension too big" errors under TeX and "off page" errors
under your driver.  

  You can dictate the edgescale for an individual figure by giving
the scale in brackets immediately after \SetLabels.  Thus, to
import into an article using say \Edge{100} a figure labelled using
another edgescale, say the original 1-by-1 default, you can use
\SetLabels[1]...\endSetLabels.

GETTING IT DOWN PAT

     Complicated labeling deserves the same respect as
complicated mathematics.  Do not expect it to come out perfect the
first time!  What is needed in either case is a mechanism to
repeatedly typeset troublesome pieces.

     One mechanism is always available.  One does complicated
labelling in a separate "test" file involving just the figure being
labelled;  a texpert will know how to \dump TeX's current state as
a temporary format that restarts rapidly at each retry.  Usually,
one then pastes the completed labelled figure back into the main
TeX file, but, of course, one can also \input it as an auxiliary
file.

     If you do not have a TeXpert at handy, here is a first
approximation to an efficient setup. By deletions reduce a copy
of your article to just a few lines before and after the figure.
Now label the figure, and finally, copy and paste the labelled
figure to the original article. Then copy the next figure to label
into this testbed and repeat. The TeXpert can improve the  speed
at which TeX starts up, by compiling a format specifically for
your article; just one caution: best NOT include in the format
ephemeral details of setup like \Set<mydriver>ArtSpecials (from
boxedeps.tex because this reads  figure dimensions which you may
change during your work session.

     An improved mechanism to repeatedly typeset troublesome
pieces is now available on the Macintosh; it is called LinoTeX;
see the same ftp sources.  It could be set up on many types
of computer.

     Before using labelfig.tex to attach labels to a graphics
object inserted using boxedeps.tex or BoxedArt.tex, make it a
firm rule to carefully adjust the bounding box using the trimming
commands of these packages, and also at least tentatively scale
and position the object. Beware of changing the grid inadvertently
after the labels have been positioned.  For example, correcting
the bounding box of a PostScript graphics object can foul up the
labels by changing the coordinate grid to which the labels are
attached. This is particularly true for the trimming  commands of
boxedeps.tex and BoxedArt.tex. However, as noted already, change
of scale is much less disruptive, and modest adjustments should be
well tolerated.

     Sometimes the labels protrude so far from the bounding box
of a figure that the figure has to be repositioned.  Best do this
by ad hoc spacing, say using \hglue and \vglue; altering the
bounding box would create a vicious circle.

     Remember that you are responsible for preventing labels
from overlapping. You are responsible for all label typography
including size and style. A label is really just about anything
that can be put in a TeX box. Note that spaces at the beginning
and end of labels will normally be suppressed; if you really want
them you must protect them with TeX braces.

     This package temporarily sets the \mathsurround parameter
of TeX to zero  while the labels are being affixed. This is done
because nonzero \mathsurround space would influence the position
of left and right aligned labels; then, when a texpert or printer
modifies mathsurround, diagram labeling might be disastrously
altered. There is a small price to pay involving labels that are
formatted as caption boxes including mathematics: you  may want or
need to specify an explicit mathsurround space within the caption
box; it will not influence anything outside.

     Those hostile to the use of * as separator between
the X and Y coordinates of label insertion points, are free to
impose another using \SetXYSeparator{<the new separator>}.  
Americans may prefer "," to "*" since they never use a 
comma as a decimal point; on the other hand, * may be more visible.

APPENDIX (I)  MERGING labelfig.tex LABELS INTO AN EPSF GRAPHICS OBJECT.

     As promised in the introduction, here is a recipe useful for
publishers. It works at least on Macintosh and at least for vectorized
graphics and Adobe type1 fonts.  (There is surely a similar recipe for
PCs under MSWindows.)

 (a)  Use boxedeps.tex utility to integrate the figure given by the eps
file, "x.eps" say, with a visible frame around it.  See
\ShowDisplacementBoxes command in boxedeps.tex.  To get precise results
automatically it is important to use the \Trim... commands of
boxedeps.tex making the "DisplacementBox" neatly fit the figure.

 (b)  Use the TeX printer driver and LaserWriter (versions >= 8.1.1) to
export to an EPSF the DVI page containing the integrated, labelled
figure. You now have an EPS file  "xx.eps"  that contains too much, and at
the wrong scale, and at wrong position.

 (c)  Convert the EPSF to an Adode Illustrator format EPSF using
the shareware utility called epsConvert by Sam Weiss
1993-- (currently $25).

 (d)  In Illustrator (or a compatible program), group the labels and the
"DisplacementBox"; copy them to the clipboard and paste them into "x.ps".
This step requires that all the label fonts be "visible to the Macintosh.

 (e)  Translate and scale the pasted group consisting of the labels plus
the "DisplacementBox" so as to make the "DisplacementBox" the bounding
box of (labelless) figure represented by "x.eps".  At this point the
labels will be correctly placed on the figure "x.eps".

 (f)  Ungroup and delete the "DisplacementBox".  The result is the
desired single EPS file, "x+.eps" say, It contains the original figure
plus its labels.  

     Using grouping and ungrouping appropriately in "x+.eps", a
publisher's staff can very efficiently improve label positions etc.

APPENDIX II)  SOME EXOTIC APPLICATIONS

     The grid of labelfig.tex is analogous to a light-table in
classical page makeup with wax or latex glue.  In principle, you
can use it to compose any page from its indivisible parts.  This
even has some of the artisanal charm of classical paste-up
provided you have a fast screen preview to make the process
"interactive".

     In practice labelfig.tex is a tool for nonstandard jobs.
Here are a few going beyond the labelling already discussed.

(I)  GRAPHICS INTEGRATION.

     This is accomplished by treating the imported graphics
objects as labels.  The underlying graphics object is then
typically an empty  \vbox to <dimension>{\vfill} in a TeX
\midinsert...\endinsert construction.  A label line
might be of the form

   (.1*.1) \\

The exact form of the special command varies from driver to
driver.  However, in the case of encapsulated PostScript graphics
(EPSF norm), by relying on boxedeps.tex, one can have the
following standard syntax (independant of driver  (see
boxedeps.doc for details.
  
  (.1*.1) \BoxedEPSF{MyFigure scaled <scale in mils>}\\

This may be slow since it requires TeX to read the PostScript
file to read bounding box using many complex macros.  So you
may want to try

  (.1*.1) \EPSFSpecial{MyFigure}{<scale in mils>}\\

which is fast and driver independant, but it squashes the
bounding box, normally to its lower left corner.

     Similarly for graphics of the Macintosh PICT norm ---
using BoxedArt.tex (same sources) in place of boxedeps.tex.

     This approach to integration is to be recommended when
one is assembling a composite graphics object.

 (II)  COMMUTATIVE DIAGRAM ENHANCEMENT

     Commutative diagrams or arrays of mathematical objects
connected by arrows of various sorts are common in mathematics.
The mathematical objects require the use of TeX.  Recently TeX
acquired a good collection of arrows of all slopes --- that of
LamSTeX --- plus pwerful macros to build the diagrams.

     However, even the LamSTeX collection is often
inadequate; it lacks for example double shafted arrows, dotted
arrows and curved arrows. Fortunately it is possible to produce
such arrows on an individual basis using sophisticated graphics
programs such as Illustrator and AldusFreehand (both serving
the EPSF norm) or using Metafont (with its public domain norm).
Since the creation of each new arrow is a work of love, you
probably want to limit the number of arrows by using LamSTeX
for most arrows. The 40K commutative diagram module of LamSTeX
has been adapted to work with AmSTeX and a copy may be posted
with LabelFig and related files. Unfortunately no one has yet
offered a version that works with Plain TeX or LaTeX.

       Suffice it here to say that when the exotic arrow has
been somehow imported into TeX, labelfig.tex treats it as a
label that one affixes to the commutative diagram.  Two other
steps will be treated in separate notes, namely the matter of
extracting the dimension specifications for the arrow and the
construction of the arrow --- for these steps are far from
unique and often depend intimately on your computer environment. 
Notes for the Macintosh-Textures-Illustrator combination are
found in the file ExoticArrows.doc.

 (III) NESTING 

Ingenuity pays off in exploiting labelfig.tex. One can
mix graphics and typography quite freely.  labelfig.tex is good
for freeform or overlapping arrangements, while boxedeps.tex (or
BoxedArt.tex) is best for regimented non-overlapping
arrangements --- and the two can be combined.

     The default behavior of labelfig.tex is not ideal 
for nesting objects, because to prevent trouble for beginners
the register for labels is globally cleared when \AffixLabels
concludes.  But there are switches available

      \LabelsGlobal      \LabelsLocal

which change this.  To understand this, extend the above test 
by something like:

 \LabelsLocal

 \SetLabels
    (.5*.5) AAA\\
 \endSetLabels

 {
 \SetLabels
    (.5*.5) ZZZ\\
 \endSetLabels
   \AffixLabels{\FirstQuadrant}
 }

   \AffixLabels{\FirstQuadrant}

     There are however potential pitfalls.  Neither
labelfig.tex nor boxedeps.tex has been tested under extreme
conditions. Problems may occur if their procedures are
indiscriminately nested. For boxedeps.tex (not labelfig.tex)
there is a precise cause for worry, namely many of its
variables are "global", which means that TeX braces will not
provide the protection one might expect.

COMMAND SUMMARY FOR labelfig.tex

  Here [...] means optional (one or zero)
       [...]* means any number of such constructs

  \SetLabels
    [[<P>](<X><Sep><Y>) <label> \\]*
  \endSetLabels
  \ShowGrid  
  \AffixLabels{<the figure>}

   --- <P> is tack position, one of eleven or empty
              order irrelevant

                   \L\T      \T      \R\T

                   \L\E      \E      \R\E

                     \L               \R

                   \L\B      \B      \R\B

   --- (<X><Sep><Y>) insertion point;
  <Sep> is separator, = * by default;
  \SetXYSeparator{<Sep>} changes it.
   <X> and <Y> are real numbers

  --- <label> a label to attach 

  --- <the figure> the figure to label 

  \GlobalLabels (default)     
  \LocalLabels  setting for nested constructs.

 \Grids makes ALL grids appear; \HideGrid then makes just next disappear.
 \noGrids returns to default.  The commands are always global.

 \GridLineWidth{<dimension>} adjusts width of grid lines. Default is very
small, to give "hairline" effect. If your grid lines are missing try
setting \GridLineWidth{1pt}.

 \Edges#1 globally changes the edge size of all grids to the numerical 
value #1, which must be 1, 10, 100, or 1000.  The default is 1.

VERSION HISTORY.
 --- Jan 1993: \Edges#1 and [??] option after \SetLabels
 --- July 1992: \Grids, \noGrids, \HideGrid;
       Gridlines become hairlines; \GridLineWidth{<dimension>}.
 --- Oct 1991, Jan 1992: \SetXYSeparator{<Sep>},  \LabelsGlobal,
       \LabelsLocal.
 --- July 1991: first release

Address for bugs and other feedback:

        Raymond S\'eroul
        IREM and Lab. de Typographie Informatise
        Univ. Rene Descartes
        Strasbourg

    Tel 33-88-41-63-45
    Email:  A18645@FRCCSC21.BITNET

        Laurent Siebenmann
        Mathematique, Bat. 425,
        Univ de Paris-Sud,
        91405-Orsay,
        France

    Tel 33-1-6941-7949; 
    Email: lcs@topo.math.u-psud.fr

\newtheorem{proposition}{Proposition}

\newtheorem{lemma}{Lemma}

\newtheorem{problem}{Problem}

\newtheorem{remark}{Remark}

\newcommand{\argmin}{\operatornamewithlimits{argmin}}
\newcommand{\argmax}{\operatornamewithlimits{argmax}}

\usepackage[dvips]{graphics}
\usepackage[dvips]{graphicx}
\usepackage{amsmath,amssymb,amsfonts,latexsym,verbatim,color,epsfig,psfrag}
\usepackage{times,multirow,multicol, array}
\usepackage{algorithm,algorithmic}
\usepackage{cite}
\usepackage{setspace}
\usepackage{url}
\usepackage{stackrel}

\definecolor{gray}{rgb}{0.5,0.5,0.5}

\begin{document}

\title{Pilot Beam Pattern Design for Channel Estimation \\ in Massive MIMO Systems}

\author{\authorblockN{Song Noh, Michael D. Zoltowski, Youngchul Sung$^\dagger$\thanks{$^\dagger$Corresponding author}, and David J. Love\thanks{
S. Noh, M. Zoltowski, and D. J. Love are with the School of Electrical and Computer Engineering, Purdue University, West Lafayette, IN 47907, USA (e-mail:songnoh@purude.edu and \{mikedz,djlove\}@ecn.purdue.edu). Y. Sung is with the Department of Electrical Engineering, KAIST, Daejeon, South Korea 305-701 (e-mail: ysung@ee.kaist.ac.kr).
This research was supported by Basic Science Research Program through the National Research Foundation of Korea (NRF) funded by the Ministry of Education (2013R1A1A2A10060852).
A preliminary version of this work was presented in
\cite{Noh&Zoltowski&Sung&Love:13ASILOMAR}, in which only the MISO
case is considered. In this paper, the sequential design proposed
in \cite{Noh&Zoltowski&Sung&Love:13ASILOMAR} is extended to the
MIMO case, power allocation, and the block-fading case. Extensive
simulation results with some realistic channel models
are provided.
}}}

\maketitle

\begin{abstract}
In this paper, the problem of pilot beam pattern design for
channel estimation in massive multiple-input multiple-output
systems with a large number of transmit antennas at the base
station is considered, and a new algorithm for pilot beam pattern
design for optimal channel estimation is proposed under the
assumption that the channel is a stationary Gauss-Markov random process.
The proposed algorithm designs the pilot beam pattern sequentially by
exploiting the properties of Kalman filtering and the associated prediction error
covariance matrices and also the channel statistics such as spatial and
temporal channel correlation.  The resulting design generates a sequentially-optimal
sequence of pilot beam patterns with low complexity for a given
set of system parameters.  Numerical results show the effectiveness
of the proposed algorithm.
\end{abstract}

\vspace{-0.3em}
\section{Introduction} \label{sec:introduction}


Multiple-input multiple-output (MIMO) systems with large-scale
transmit antenna arrays, so called {\em massive MIMO} systems, is
one of the key technologies for future wireless communications.
The large size of the transmit antenna array relative to the
number of receive terminals can average out thermal noise,  fast channel
fading, and some interference, based on  the law of large
numbers\cite{Marzetta:10WCOM,Rusek&Persson&Lau&Larsson&Edfors&Tufvesson&Marzetta:13SPM}.
Massive MIMO provides high data rates and energy efficiency with
simple signal processing because the propagation channels to
terminal stations served by a base station equipped with massive
MIMO are asymptotically orthogonal due to the increased beam
resolution
\cite{Shepard&Yu&Anand&Li&Marzetta&Yang&Zhong:12MobiCom,Gao&Tufvesson&Edfors&Rusek:12ASILOMAR}.
However, in practice, such benefits may be limited by channel estimation
accuracy
\cite{Jose&Ashikhmin&Marzetta&Vishwanath:11WCOM}. This is especially
true when full frequency reuse across neighboring cells is adopted;  in this case,
pilot contamination
\cite{Marzetta:10WCOM,Jose&Ashikhmin&Marzetta&Vishwanath:11WCOM,
Yin&Gesbert&Filippou&Liu:13JSAC,Ngo&Larsson:12ICASSP}
leads to imperfect channel estimation which, in turn, yields severely degraded
system performance.
Furthermore, in contrast to the conventional MIMO system employing
a small number of antennas, the overhead required for channel
estimation for massive MIMO can be overwhelming and thereby severely limit
the above mentioned benefits of massive MIMO. Since the available
training resources are limited by either the channel coherence
interval or the amount of interference induced by neighboring
cells, fast and reliable channel estimation with reduced
training overhead is critical to massive MIMO systems.

To tackle the challenge of channel estimation, much of the prior work
focused on time-division duplex (TDD) operation assumed
channel reciprocity
\cite{Marzetta:10WCOM,Rusek&Persson&Lau&Larsson&Edfors&Tufvesson&Marzetta:13SPM,Hoydis&Brink&Debbah:13JSAC},
and reciprocity
calibration\cite{Shepard&Yu&Anand&Li&Marzetta&Yang&Zhong:12MobiCom}
under the assumption of time-invariant channels within the coherence time.
More recently, Wiener prediction has been employed to mitigate the impact of channel aging
over time under the assumption of time-varying channels
\cite{Truong&Heath:13JCN}.  However, in most wireless systems,
frequency-division duplex (FDD) operation is employed, and in
this case the problem of channel estimation becomes more
challenging because MIMO channel sounding requires substantial
overhead (such as feedback and/or dedicated times for channel sounding)
that scales with the number of antennas.  Such
overhead can limit the performance improvement that is expected in
massive MIMO systems. There has been some work on channel
estimation and channel state information (CSI) feedback techniques
for FDD massive MIMO systems, based on compressive sensing
\cite{Kuo&Kung&Ting:12WCNC}, limited feedback
\cite{Choi&Chance&Love&Madhow:13ITA,Choi&Love&Madhow:13CISS}, and
projected channels \cite{Nam&Ahn&Adhikary&Caire:12CISS}. Also, to
improve channel estimation performance, the problem of pilot
beam design was investigated for massive MIMO systems under the
assumption of closed-loop training
\cite{Kudo&Armour&McGeehan&Mizoguchi:12ISWCS,Love&Choi&Bidigare:13CISS}.

In this paper, we consider the problem of pilot beam design for
downlink channel estimation in FDD massive MIMO systems, for the case where the
number of symbol times for channel sounding within a channel
coherence time is typically much less than the number of antennas.
To design efficient pilot beam patterns, we here exploit
channel statistics for massive MIMO systems derived from {\em dynamic channel
modelling}\cite{Baddour&Beaulieu:05WCOM,Tong&Sadler&Dong:04SPM,Yu&Sung&Kim&Lee:12SP}
and analytical {\em channel spatial
correlation} models \cite{Sayeed:02SP,Shiu&Foschini&Gans&Kahn:00COM,Forenza&Love&Heath:07VT,Adhikary&Nam&Ahn&Caire:13IT}.
{Since the gain of beamforming in practical wireless systems is obtained mainly in slowly fading channels, we focus on slowly fading and exploit the correlated time-variations
in the channel by adopting the widely-used Gauss-Markov channel model \cite{Dong&Tong&Sadler:04SP}.}
Under this model, the
channel estimation performance can be enhanced through the use of optimal
Kalman filtering and prediction that exploits the current and all
previously received pilot signals, thereby shortening the required time for
accurate channel estimation.   Our model also
incorporates spatial channel correlation that depends on both the
antenna geometry and the scattering environment;
experimental investigations and analytical studies have confirmed that
this information is typically available in (massive) MIMO systems
\cite{Shepard&Yu&Anand&Li&Marzetta&Yang&Zhong:12MobiCom,Gao&Tufvesson&Edfors&Rusek:12ASILOMAR,Sayeed:02SP,Shiu&Foschini&Gans&Kahn:00COM,Forenza&Love&Heath:07VT,Adhikary&Nam&Ahn&Caire:13IT}
and is locally\footnote{It means that for a short period of
time, the correlation characteristics do not change much.}
time-wise stationary \cite{Stein:87JSAC}.  By exploiting both the
channel dynamics and the spatial correlation, we develop a low-complexity
pilot beam pattern design procedure that provides a sequence of
optimal pilot beam patterns that sequentially minimize the channel estimation
mean square error (MSE) at each training instant based on a greedy
approach. (The definition of sequential optimality will be
provided soon.)  The key idea underlying the proposed method is the
joint use of spatio-temporal channel correlation and
signal-to-noise ratio (SNR) combined with the exploitation of the
structure of the error covariance matrices generated with optimal
Kalman filtering under the Gauss-Markov model, to derive a sequence
of optimal pilot beam patterns for each training period.

This paper is organized as follows: The system model and
background are described in Section \ref{sec:systemmodel}. Section
\ref{sec:proposedmethod} describes  the proposed pilot beam
pattern design method. Practical issues of implementing the proposed method are discussed in Section \ref{sec:discussion}. 
Numerical results are provided in Section
\ref{sec:numericalresult}, followed by conclusions in Section
\ref{sec:conclusion}.


\noindent  \textbf{Notation}  Vectors and matrices are written in
boldface with matrices in capitals. All vectors are column
vectors. For a matrix $\Abf$,  $\mathbf{A}^T$, $\mathbf{A}^H$, and
$\mathbf{A}^{\ast}$ indicate the transpose, Hermitian transpose,
and  complex conjugate of $\mathbf{A}$, respectively.
$\mbox{tr}(\mathbf{A})$ and $\text{var}(\mathbf{A})$ denote the
trace of $\mathbf{A}$ and the variance operator, respectively.
$\text{vec}(\mathbf{A})$ denotes the column vector obtained by
stacking the elements of $\mathbf{A}$ columnwise.
$[\mathbf{A}]_{i,j}$ denotes the element of $\mathbf{A}$ at the
$i$-th row, and $j$-th column.  $\text{diag}(a_1,\cdots,a_n)$
denotes a diagonal matrix with diagonal elements $a_1,\cdots,a_n$,
whereas $\text{diag}(\mathbf{A})$ is the column vector containing
the diagonal elements of a matrix $\mathbf{A}$. For a vector
$\abf$, we use $\|\abf\|_1$ for $1$-norm and $\|\abf\|_2$ for
$2$-norm. For two matrices $\Abf$ and $\Bbf$, $\Abf \otimes \Bbf$
denotes the Kronecker product, and $\Abf \preceq \Bbf$ means that
$\Bbf-\Abf$ is positive semi-definite.  $E\{\mathbf{x}\}$
represents the expectation of $\mathbf{x}$. $\mathbf{I}_n$ stands
for the identity matrix of size $n$, and ${\mathbf{1}}$ denotes a
column vector with all one elements. ${\mathbb{R}}_+$ denotes the
set of non-negative real numbers. $\iota = \sqrt{-1}$ is used for the imaginary
number so that $i$ and $j$ may be used as indices.

\vspace{-0.4em}
\section{System Model}\label{sec:systemmodel}
\vspace{-0.1em}

\subsection{System Setup}

\begin{figure}[!t]
\centerline{
\psfrag{(nt1)}[c]{\footnotesize $1$} %
\psfrag{(nt2)}[c]{\footnotesize {$2$}} %
\psfrag{(nt)}[c]{\footnotesize {$N_t$}} %
\psfrag{(sk)}[c]{\footnotesize $d_k$} %
\psfrag{(ck)}[c]{\footnotesize $\Sbf_k$} %
\psfrag{(md)}[c]{\scriptsize $(k \notin \mathcal{I}_p)$} %
\psfrag{(mp)}[c]{\scriptsize $(k \in \mathcal{I}_p)$} %
\psfrag{(nr1)}[c]{\footnotesize $1$} %
\psfrag{(nr)}[c]{\footnotesize {$N_r$}} %
\psfrag{(yk)}[c]{\footnotesize {$\ybf_k$}} %
\psfrag{(hest)}[c]{\footnotesize {$\hat{\hbf}_{k|k}$}} %
\psfrag{(h)}[c]{\footnotesize {$\mathbf{H}_k$}} %
\psfrag{(kalman)}[c]{\scriptsize MMSE} %
\psfrag{(filter)}[c]{\scriptsize filter} %
\psfrag{(proposed)}[c]{\scriptsize Tracking $\boldsymbol\lambda$} %
\includegraphics[scale=0.86]{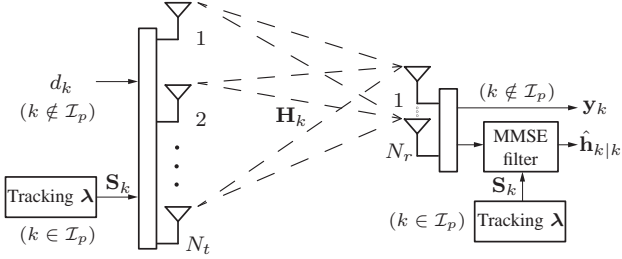}}
\vspace{-0.5em}
\caption{Massive MIMO system model where $\lambdabf$ is the eigenvalues of the prediction covariance matrix $\Pbf_{k|k-1}$} \label{fig:massiveMIMOsystemModel}
\vspace{-1.3em}
\end{figure}

We consider a massive MIMO system with $N_t$ transmit antennas and
$N_r$ received antennas $(N_t \gg N_r)$, where the channel is
given by an $N_r\times N_t$ MIMO system with flat Rayleigh fading
under the narrowband assumption
\cite{Molisch&Foerster&Pendergrass:03WCOM} (which easily extends
to the case of wideband frequency-selective channel when the
system adopts OFDM transmission\cite{Jakes:book}).  The received
signal at the $k$-th symbol time is given by
\begin{align}
\ybf_k  &= \Hbf_k\sbf_k^* +
\wbf_k,~~~k=1,2,\ldots\label{eq:statespacemodel_y1}
\end{align}
where $\sbf_k$ is the $N_t\times 1$ transmitted symbol vector at
time $k$, $\Hbf_k$ is the $N_r\times N_t$ MIMO channel matrix at
time $k$, and $\wbf_k$ is the zero-mean independent and
identically distributed (i.i.d.) complex Gaussian noise vector at
time $k$ with covariance matrix  $\sigma_w^2\Ibf_{N_r}$, as shown
in Fig. \ref{fig:massiveMIMOsystemModel}.
{(Here, we used the complex conjugate on $\sbf_k$ to keep the notation consistent with \eqref{eq:statespacemodel_y2}.)}

\subsubsection{MIMO Channel Correlation Model}

For channel correlation, we consider the general Kronecker model
that captures the transmit and receive antenna
correlation\cite{Shiu&Foschini&Gans&Kahn:00COM,Forenza&Love&Heath:07VT}.
The transmit and receive channel covariance matrices reflect the
geometry of the propagation paths and {remain almost unchanged (locally time-wise) when compared to the rapidly-varying instant channel
realization,} since the array response to the scattering
environments changes slowly compared to the user's
location\cite{Stein:87JSAC,Gerlach&Paulraj:94GLOBECOM}. Thus, the
channel covariance matrices are assumed to be fixed over the
considered time period for channel estimation, and the considered
Kronecker channel  model is given by
\begin{align}
\Hbf_k &= \Rbf_r^{1/2} \tilde{\Hbf}_k (\Rbf_t^{1/2})^T,
\label{eq:mimoChannelModel_v2}
\end{align}
where $\{\tilde{\Hbf}_k \in \mathbb{C}^{N_r\times N_t},
k=1,2,\cdots\}$ is an ergodic sequence of  random matrices with
independent zero-mean Gaussian elements with some variance, and
$\Rbf_t\in\mathbb{C}^{N_t\times N_t}$ and
$\Rbf_r\in\mathbb{C}^{N_r\times N_r}$ are deterministic transmit
and receive correlation matrices, respectively, i.e.,
$\Rbf_t=\frac{1}{N_r}E\{\Hbf^H_k\Hbf_k\}$ and
$\Rbf_r=\frac{1}{N_t}E\{\Hbf_k\Hbf^H_k\}$ so that
$\text{tr}(E\{\Hbf_k\Hbf_k^H\})=N_tN_r$. (Case studies for some
channel models are discussed in\cite{Tulino&Lozano&Verdu:05IT}.)

In the downlink training, the channel covariance matrices can be
estimated by subspace estimation methods even without the
knowledge of instantaneous channel state information
\cite{Eriksson&Stoica&Soderstrom:94SP,Tong&Perreau:98IEEE,Noh&Sung&Zoltowski:13WCOM},
and {there also exist methods that estimate the downlink channel covariance
matrix using uplink training in FDD systems using techniques such as
frequency calibration matrix\cite{Liang&Chin:01JSAC}, log-periodic
array\cite{Hochwald&Marzetta:01SP}, or duplex array
approach\cite{Raleigh&Diggavi&Jones&Paulraj:95ICC}.} Furthermore,
under some circumstances the channel covariance matrices $\Rbf_t$
and $\Rbf_r$ are approximately known {\em a priori}. For example, under the
virtual channel condition \cite{Sayeed:02SP}, the use of uniform
linear arrays (ULAs) at the transmitter and the receiver makes
$\Rbf_t$ and $\Rbf_r$ approximately Toeplitz. By extending the
{\em one-ring} model introduced by Jakes \cite{Jakes:book}, the
spatial correlation in the flat-fading case
 can be determined by the physical environment such as angle
spread (AS), angle of arrival (AoA), and antenna
geometry\cite{Shiu&Foschini&Gans&Kahn:00COM}. That is, in the case
of a ULA with the AoA $\theta$ and the antenna spacing $\lambda
D$, the channel covariance matrix is given by
\vspace{-0.1em}
\begin{align}
[\Rbf_t]_{i,j} &= \frac{1}{2\Delta}
\int^{\theta+\Delta}_{\theta-\Delta} e^{-\iota 2\pi D(i-j)
\sin(\alpha)} d\alpha, \label{eq:channelCovOneRing}
\end{align}
where $\lambda$ is the wavelength and $\Delta$ is the AS. (This
result can be  extended to two-dimensional or planar
arrays\cite{Adhikary&Nam&Ahn&Caire:13IT}.) When the number of
transmit antennas grows large, the eigenspace of $\Rbf_t$ is
closely approximated by a unitary {\em Discrete Fourier Transform}
(DFT) matrix with the support of AoA distribution. Hereafter, we
shall assume that the transmitter and the receiver have the
knowledge of the channel covariance matrices.
{The assumption of known $\Rbf_t$ will be revisited in Section \ref{sec:discussion}.}

\subsubsection{Channel Variation in Time and Slotted Transmission
Structure}

For channel variation in time, we adopt a state-space model, i.e.,
the channel dynamic is given by the first-order stationary
Gauss-Markov process
\cite{Iltis:90COM,Baddour&Beaulieu:05WCOM,Tong&Sadler&Dong:04SPM,Yu&Sung&Kim&Lee:12SP}
\begin{equation}
\hbf_{k+1}  = a \hbf_k + \sqrt{1-a^2}\bbf_k
\label{eq:statespacemodel_h}
\end{equation}
that satisfies the Lyapunov equation
\begin{equation}
\Rbf_\hbf = a^2 \Rbf_\hbf + (1-a^2) \Rbf_\bbf , \label{eq:lyapunov}
\end{equation}
and $\Rbf_\hbf =E\{\hbf_k \hbf_k^H\}=\Rbf_\bbf = E\{\bbf_k
\bbf_k^H\}$ for all $k$ \cite{Stein:87JSAC}, where
$\hbf_k:=\text{vec}(\Hbf_k)$, $\bbf_k$ is a zero-mean and
temporally independent plant Gaussian vector, and $a \in (0,1]$ is
the temporal fading coefficient.\footnote{For Jakes' model,
$a=J_0(2\pi f_D T_s)$\cite{Jakes:book}, where $J_0(\cdot)$ is the
zeroth-order Bessel function, $T_s$ is the transmit symbol
interval, and $f_D$ is the maximum Doppler frequency shift.} (It
is easy to verify that $\{\hbf_k,k=1,2,\cdots\}$ is a stationary
process under this assumption.)  The temporal fading correlation
coefficient $a$ can be estimated
\cite{Iltis:90COM,Tsatsanis&Giannakis&Zhou:96ICASSP,Dai&Zhang&Xu&Mitchell&Yang:12EURASIP,Liu&Hansson&Vandenberghe:13SCL},
and  we assume that $a$ is known. Then, under the Kronecker
channel model \eqref{eq:mimoChannelModel_v2} we have
\begin{equation}
\Rbf_{\hbf}=\Rbf_t\otimes \Rbf_r.
\end{equation}

We assume  slotted transmission  with $M$ consecutive symbols as
one slot which is comprised of a training period of $M_p$ symbols
and a data transmission period of $M_d$ symbols so that
$M=M_p+M_d$.

\subsection{Channel Estimation}\label{subsec:channelestimation}

We consider the minimum mean square error (MMSE) approach for
channel estimation \cite{sungetal:06book} based on the current and
all previous observations during training periods, i.e.,
$\hat{\hbf}_{k|k}:=E\{\hbf_k|\ybf_p^{(k)}\}$ where $\ybf_p^{(k)}$
denotes all received signals during the pilot transmission up to
symbol time $k$, given by
\begin{align}
\ybf_p^{(k)}&=\{\ybf_{k^\prime}| k^\prime \le k, k^\prime \in \Ic_p  \},\nonumber
\end{align}
where $\Ic_p:=\{ k=l M+m | l =0,1,2,\cdots, m=1,\cdots,M_p\}$. At
each training symbol time, a pilot beam vector (or beam pattern)
$\sbf_k$ of size $N_t$, $k \in \Ic_p$,  is transmitted for channel
estimation.  During the data transmission period, on the other
hand, the base station sends unknown data with transmit
beamforming based on the estimated channel.\footnote{
{Transmit beamforming in FDD requires feedback information for channel state
information (CSI) from the receiver. Thus, the quantized version
of the downlink channel or the index of the quantized version of the channel chosen from a receiver can be fed back to the base station\cite{Love&Heath&Santipach&Honig:04COMMAG}.
In addition, a quantized (or analog) version of the received training signal $\ybf_k\in\mathbb{C}^{N_r}$ can be fed back to enable channel estimation at the base station\cite{Noh&Zoltowski&Sung&Love:13ASILOMAR}.
The focus of the paper is not feedback quantization but optimal design of the
pilot beam pattern for channel estimation.}}

Note that the received signal model \eqref{eq:statespacemodel_y1}
can be rewritten as
\begin{align}
\ybf_k  &= \Sbf_k^H\hbf_k + \wbf_k,\label{eq:statespacemodel_y2}
\end{align}
where $\Sbf_k:= \sbf_k \otimes \Ibf_{N_r}$ is an $N_tN_r \times
N_r$ matrix. Then, we have a state-space model obtained from
\eqref{eq:statespacemodel_h} and \eqref{eq:statespacemodel_y2} and
the optimal channel estimation is given by the Kalman filter for
this state-space model\cite{Kailath:book}.  During  the training
period, the Kalman filter performs a measurement update step for
channel estimation at each symbol time, where the Kalman channel
estimate and the related error covariance matrices are given
by\cite{Kailath:book}
\begin{align}
\hat{\hbf}_{k|k} &= \hat{\hbf}_{k|k-1}+\Kbf_k(\ybf_k-\Sbf_k^H\hat{\hbf}_{k|k-1})\label{eq:measurementupdateH}\\
\Pbf_{k|k-1} &= a^2\Pbf_{k-1|k-1} + (1-a^2)\Rbf_{\hbf},\label{eq:mesurementupdateP2}\\
\Pbf_{k|k} &= \Pbf_{k|k-1} - \Kbf_k\Sbf_k^H\Pbf_{k|k-1}, \label{eq:mesurementupdateP}
\end{align}
where
$\Kbf_k=\Pbf_{k|k-1}\Sbf_k(\Sbf_k^H\Pbf_{k|k-1}\Sbf_k+\sigma_w^2\Ibf_{N_r})^{-1}$,
$\hat{\hbf}_{1|0}={\mathbf{0}}$, and $\Pbf_{1|0}=\Rbf_\hbf$. Here,
$\Pbf_{k|k}$ and $\Pbf_{k|k-1}$ are the estimation and prediction
error covariance matrices, respectively, defined as
$\Pbf_{k|k^\prime} = E\bigl\{(\hbf_k-\hat{\hbf}_{k|k^\prime})(\hbf_k-\hat{\hbf}_{k|k^\prime})^H |\ybf_p^{(k^\prime)}\bigr\},
$ where
$\hat{\hbf}_{k|k^\prime}:=E\bigl\{\hbf_k|\ybf_p^{(k^\prime)}\bigr\}$. During
the data transmission period, the channel is predicted based on
the last channel estimate of the previous training period
as\cite{Kailath:book}
\begin{align}
&\hat{\hbf}_{l M+M_p+m | l M+M_p}=a^{m}\hat{\hbf}_{l M+M_p+m|l M+M_p} \label{eq:timeupdateH}\\
&\Pbf_{l M+M_p+m|l M+M_p} =a^{2m}\Pbf_{l M+M_p|l
M+M_p}+(1-a^{2m})\Rbf_{\hbf},\nonumber
\end{align}
where  $m=1,\ldots,M_d$.  During the data transmission period, the
predicted channel can be used for  transmit beamforming; for
example,  eigen-beamforming \cite{Telatar:99,Marzetta:10WCOM}
based on the predicted channel can be applied for maximum rate
transmission.

In the simple case of multiple-input single-output
(MISO) transmission, maximal ratio transmit beamforming based on
the current channel estimate can be applied, and the transmit
signal vector in this case is given by $
\sbf_k=\frac{\hat{\hbf}_{k | l M+M_p}}{\|\hat{\hbf}_{k | l
M+M_p}\|_2} d_k$,  where $d_k$ is the data symbol at symbol time
$k$, $k=l M+M_p+m$.
{From \eqref{eq:statespacemodel_y2} and $\Delta\hbf_k:= \hbf_k-\hat{\hbf}_{k|lM+M_p}$, the received signal model can be rewritten as
\begin{align}
y_k&=\sbf_k^H\hat{\hbf}_{k|lM+M_p} + \sbf_k^H\Delta\hbf_k +w_k \label{eq:statespacemodel_y3}
\end{align}
The second term in \eqref{eq:statespacemodel_y3} denotes the additional noise resulting from imperfect channel estimation.
By using the deterministic approximation of
$\frac{1}{N_t}|\sbf_k^H \Delta\hbf_k|^2 -
\frac{1}{N_t}\sbf_k^H\Pbf_{k|lM+M_p}\sbf_k
\stackrel[N_t\rightarrow \infty]{~~\text{a.s.}~~}{\longrightarrow} 0
$ \cite{Hoydis&Brink&Debbah:13JSAC}, the received SNR with the estimated channel is defined as
\begin{align} \label{eq:receivedSNRrevision}
\text{Received SNR}
= \frac{|\sbf_k^H\hat{\hbf}_{k|lM+M_p}|^2}{\sbf_k^H\Pbf_{k|lM+M_p}\sbf_k + \sigma_w^2}.
\end{align}
}

\vspace{-0.3em}
\section{The Proposed Pilot Beam Pattern Design} \label{sec:proposedmethod}

In this section, we present our proposed pilot beam pattern design
methods that minimize the channel estimation MSE associated with
optimal Kalman filtering  explained in the previous section. The
channel estimation MSE is directly related to the effective
SNR\cite{Hassibi&Hochwald:03IT} and thus such pilot beam pattern
design can be leveraged to improve the training-based channel
capacity.

\subsection{Greedy Sequential Design} \label{subsec:ProposedMethod_1}

We notice from \eqref{eq:timeupdateH} that the channel estimation
error during the data transmission period depends only  on $a$,
$\Rbf_\hbf$ and the estimation error covariance matrix $\Pbf_{l
M+M_p|l M+M_p}$  at the last pilot symbol time.  $a$ and
$\Rbf_\hbf$ are given, but the estimation MSE at the last pilot
symbol time, $\mbox{tr}(\Pbf_{l M+M_p|l M+M_p})$, can be minimized
by properly designing the pilot beam pattern sequence $\{\sbf_k|
k={l}^\prime M+m$, ${l}^\prime \le l $, $m=1,\ldots,M_p\}$. Here,
since $\Pbf_{l M+M_p|l M+M_p}$ is a function of $\Sc:=\{\sbf_j | j
= {l}^\prime M + m, m=1,\cdots,M_p, j \le l M+M_p \}$, $\Sc$
should be jointly optimized to minimize the MSE at time $k=l
M+M_p$. However, this joint optimization is  too complicated
because the impact of $\Sc$ on $\Pbf_{l M+M_p|l M+M_p}$ is
intertwined over time.\footnote{The difficulty in applying
standard dynamic programming (DP)\cite{Bertsekas:05book} to the
problem is that the contribution of $\sbf_k$ at time $k$ to the
cost function is not localized at time $k$. It affects the
so-called branch metric at time $k$ and all the following branch
metrics.} Furthermore, optimal channel estimation at $k=lM+M_p$
for some $l$ is not the only optimization goal since the MSE at
$k=l^\prime M+M_p$ for each and every $l^\prime$ should be
optimized for the $l^\prime$-th data transmission period.
Therefore, we first adopt a greedy sequential optimization
approach to design the pilot beam pattern sequence, which is
formally stated as follows.

\vspace{0.5em}
\begin{problem} \label{prob:problemstatement}
For each pilot symbol time $k$ starting from 1, given $\sbf_{j}$
for all pilot symbol time $j < k$, design $\sbf_k$ such that
\begin{align}
\min_{\sbf_k}&~\text{tr}\left(\Pbf_{k|k}\right)\label{eq:objftntracePkk}\\
\text{s.t.}&~\|\Sbf_k\|_F^2=N_r\|\mathbf{s}_k\|_2^2=N_r\rho_p.
\end{align}
\end{problem}
The solution to Problem \ref{prob:problemstatement} is given by the following proposition.

\vspace{0.5em}
\begin{proposition}
\label{pro:argminMMSE_mimo} Given all previous pilot signals
$\sbf_{j}$ ($j<k$),

{\em i)} in the MISO case, the pilot beam pattern $\sbf_k$ at time
$k$ minimizing $\text{tr}(\Pbf_{k|k})$ is given by a scaled
dominant  eigenvector of the error covariance matrix
$\Pbf_{k|k^\prime}$ of the Kalman prediction  for time $k$
\cite{Noh&Zoltowski&Sung&Love:13ASILOMAR}, and

{\em ii)} {in the MIMO case, if
the Kalman prediction error covariance matrix
$\Pbf_{k|k^\prime}$ for time $k$ is decomposed as
\begin{equation}   \label{eq:MIMOecmdecomp}
\Pbf_{k|k^\prime}=(\Ubf\otimes\Vbf)\mbox{diag}(\Lambdabf_1,\cdots,\Lambdabf_{N_t})(\Ubf\otimes\Vbf)^H,
\end{equation}
where $\Ubf\in\mathbb{C}^{N_t \times N_t}$ and $\Vbf
\in\mathbb{C}^{N_r \times N_r}$ are unitary matrices, and
$\Lambdabf_i \in\mathbb{R}_+^{N_r \times N_r}$ is a diagonal
matrix with nonnegative real elements,\footnote{This assumption will be verified shortly in Proposition
\ref{pro:periodicityPilotPattern_mimo}.}
then a locally optimal  pilot beam pattern
$\sbf_k$ at time $k$ for minimizing $\text{tr}(\Pbf_{k|k})$ is
given by a scaled version of a column vector of the unitary matrix
$\Ubf$ in \eqref{eq:MIMOecmdecomp}.}
\end{proposition}

\vspace{0.5em}
{\em Proof:} See Appendix \ref{app:argminMMSE_mimo}.

\vspace{0.5em}

Interestingly, it can be shown in the MISO case that
 the pilot beam pattern $\sbf_k$ obtained from \eqref{eq:objftntracePkkMIMO_v2}  is equivalent to the first  principal component direction of $\Pbf_{k|k-1}$ given by
\begin{align}
{\mathop{\arg\max}}_{\|\sbf_k\|_2^2=\rho_p}&~\text{var}\left(\sbf_k^H(\mathbf{h}_k-\hat{\mathbf{h}}_{k|k})\right).
\end{align}

As seen in the proof, in the MIMO case, it is not easy to obtain a
globally optimal solution, but the obtained locally optimal
solution yields  a nice property that can be exploited to derive
an efficient pilot beam pattern design algorithm.  Note that to
obtain the (sequentially)  optimal $\sbf_k$, we need to perform
the eigen-decomposition (ED) of $\Pbf_{k|k^\prime}$  at each pilot
symbol time $k$, and this can be computationally expensive since
$N_t$ is large for massive MIMO systems.  However, due to the
following proposition regarding the eigen-space of the Kalman
prediction error covariance matrix associated with Proposition
\ref{pro:argminMMSE_mimo}, we can eliminate such heavy complexity
burden when designing a sequentially optimal pilot beam pattern
sequence.

\vspace{0.5em}
\begin{proposition}
\label{pro:periodicityPilotPattern_mimo}  The Kalman filtering
error covariance matrix $\Pbf_{k|k}$ and the Kalman prediction
error covariance matrix $\Pbf_{k|k^\prime}$ generated by
sequentially optimal $\sbf_k$ given by Proposition
\ref{pro:argminMMSE_mimo} are simultaneously diagonalizable with
$\Rbf_\hbf$ for any $k$ and $k^\prime(<k)$, under the assumption
of $\Pbf_{1|0}=\Rbf_{\hbf}=\Rbf_t\otimes\Rbf_r$.\footnote{Such an
initial parameter is a typical value for the Kalman filter, and
there will be no loss\cite{Bertsekas:05book}.}
\end{proposition}
\vspace{0.5em}

{\em Proof:} Proof is by induction. Let
$\Rbf_t=\Ubf\Sigmabf\Ubf^H$ and $\Rbf_r=\Vbf\Gammabf\Vbf^H$ be the
ED of $\Rbf_t$ and $\Rbf_r$, respectively. Then, $\Pbf_{1|0}=(\Ubf
\otimes \Vbf)\Lambdabf^{(1)}(\Ubf\otimes\Vbf)^H$, where
$\Lambdabf^{(1)}=\Sigmabf \otimes\Gammabf $.

For any pilot symbol time $k=l M+m$ ($m=1,\ldots,M_p$), suppose
that the Kalman prediction matrix for time $k$ is given by
$\Pbf_{k|k-1}=(\Ubf\otimes\Vbf)\Lambdabf^{(k)}(\Ubf\otimes\Vbf)^H$,
where $\Ubf\in\mathbb{C}^{N_t \times N_t}$ and $\Vbf
\in\mathbb{C}^{N_r \times N_r}$ are unitary matrices, and
$\Lambdabf^{(k)}\in\mathbb{R}^{N_t N_r \times N_t N_r}$ is a
diagonal matrix given as
\begin{equation} \label{eq:partitionedblkdiagLambda}
\Lambdabf^{(k)}=\mbox{diag}(\Lambdabf_{1}^{(k)}, \cdots,
\Lambdabf_{N_t}^{(k)}).
\end{equation}
By Proposition \ref{pro:argminMMSE_mimo},  $\sbf_k$ is given by a
scaled version of a column vector $\ubf_{i_k}$ of $\Ubf$, i.e.,
$\sbf_k=\sqrt{\rho_p}\ubf_{i_k}$ with
\begin{align}
i_k:=\argmax_{i}&~
\text{tr}\left\{(\rho_p\Lambdabf_i^{(k)}+\sigma_w^2\Ibf_{N_r})^{-1}
\rho_p (\Lambdabf_i^{(k)})^2\right\}.\label{eq:k_mdef_v1}
\end{align}
 Then, from the measurement update \eqref{eq:mesurementupdateP},
$\Pbf_{k|k}$ is given by
\begin{align}
\Pbf_{k|k}&=
(\Ubf\otimes\Vbf)\left\{ \Lambdabf^{(k)} - (\ebf_{i_k}\ebf_{i_k}^T)\otimes \right.\nonumber\\
&~~~\left.
\left[\rho_p\Lambdabf_{i_k}^{(k)}\bigl(\rho_p\Lambdabf_{i_k}^{(k)}+\sigma_w^2\Ibf_{N_r}\bigr)^{-1}
\Lambdabf_{i_k}^{(k)}\right]\right\}(\Ubf\otimes\Vbf)^H \nonumber\\
&=: (\Ubf\otimes\Vbf)\bar{\Lambdabf}^{(k)}(\Ubf\otimes\Vbf)^H,
\label{eq:measurementUpdateReductionMIMO}
\end{align}
where $\bar{\Lambdabf}^{(k)}$ is a diagonal matrix with
nonnegative elements. (See Appendix
\ref{subsec:equations_for_P_kk} for details.) Thus, $\Pbf_{k|k}$
and $\Pbf_{k|k-1}$ are simultaneously diagonalizable. Since
$\Rbf_{\hbf}=\Rbf_t\otimes\Rbf_r=(\Ubf\otimes\Vbf)\Lambdabf^{(1)}(\Ubf\otimes\Vbf)^H$,
$\Pbf_{k+1|k}$ from the prediction step
\eqref{eq:mesurementupdateP2}  is also simultaneously
diagonalizable with $\Rbf_{\hbf}$ since $\Pbf_{k|k}$ is
simultaneously diagonalizable with $\Rbf_\hbf$.

Now consider a symbol time $k$ during the first data transmission period. In this case, the prediction error covariance matrix is given by
\begin{align}
&\Pbf_{M_p+m|M_p} \nonumber\\
&=a^{2m}\Pbf_{M_p|M_p}+(1-a^{2m})\Rbf_{\hbf} \label{eq:datammseMIMO} \\
&=(\Ubf\otimes\Vbf) \bigl(\Lambdabf^{(1)} - a^{2m}(\Lambdabf^{(1)}
- \bar{\Lambdabf}^{(M_p)})\bigr) (\Ubf\otimes\Vbf)^H, \nonumber
\end{align}
where $m=1,\ldots,M_d$ and $\bar{\Lambdabf}^{(M_p)}$ is defined in
\eqref{eq:measurementUpdateReductionMIMO}. Thus, any prediction
error covariance matrix during the first data period is
simultaneously diagonalizable with $\Pbf_{k|k}$ for $k\le M_p$.
Since this Kalman recursion repeats, we have the claim.
$\hfill{\blacksquare}$ \vspace{0.5em}

Note that the assumption \eqref{eq:MIMOecmdecomp} is valid under
the Kronecker channel correlation model together with the pilot
beam pattern selection proposed in Proposition
\ref{pro:argminMMSE_mimo}. Proposition
\ref{pro:periodicityPilotPattern_mimo} states that all Kalman
error covariance matrices under the sequentially optimal pilot
beam pattern design have the same set of eigenvectors as
$\Rbf_\hbf$. This has an important practical implication: in each
pilot transmission period, the base station transmits a pilot beam
pattern at time $k$ chosen from a fixed set of  orthogonal beam
patterns, i.e., the transmit eigenvectors of $\Rbf_\hbf$,
according to some order depending on $\{\Lambdabf_{i}^{(k)},
i=1,\cdots,N_t\}$ (defined in
\eqref{eq:partitionedblkdiagLambda}). Note that
\eqref{eq:measurementUpdateReductionMIMO} shows how a sequentially
optimal pilot beam pattern at time $k$ reduces the channel
estimation error by changing the eigenvalue distribution from
$\Lambdabf^{(k)}$ to $\bar{\Lambdabf}^{(k)}$ with the measurement
update step (only the $i_k$-th subblock is updated as
$\bar{\Lambdabf}_{i_k}^{(k)}=\sigma_w^2(\rho_p
\Lambdabf_{i_k}^{(k)}+\sigma_w^2\Ibf_{N_r})^{-1}\Lambdabf_{i_k}^{(k)}$),
and \eqref{eq:datammseMIMO} shows how the eigenvalues of the
channel prediction error covariance matrix change (from
$\bar{\Lambdabf}^{(k)}$ to $\Lambdabf^{(k+m)}$) during the pure
prediction period. Exploiting these facts, we propose an efficient
algorithm to obtain the sequence of sequentially optimal pilot
beam patterns to minimize the channel estimation MSE at each
symbol time. The algorithm is summarized in Algorithm
\ref{alg:optimalMIMO}.

\begin{algorithm} [h]                       
\caption{Sequentially Optimal Pilot Beam Pattern Design}          
\label{alg:optimalMIMO}                 
\begin{algorithmic}                         
\REQUIRE Perform the ED of $\Rbf_t=\Ubf\Sigmabf\Ubf^H$ and
$\Rbf_r=\Vbf\Gammabf\Vbf^H$, and $\Rbf_\hbf=\Rbf_t\otimes\Rbf_r$.
Store $\lambdabf^{(1)}=\text{diag}(\Sigmabf\otimes \Gammabf)$, and
$\Ubf=[\ubf_1,\cdots,\ubf_{N_t}]$.

\STATE $\lambdabf=\lambdabf^{(1)}$ and partition
$\lambdabf=[\lambdabf_1^T,\cdots,\lambdabf_{N_t}^T]^T$

\WHILE{$l =0,1,\cdots$}

\FOR{$m = 1$ to $M$}

\STATE $k=lM+m$

\IF{$m\le M_p$}

\STATE $i_k=\argmax_i
\sum_{j=1}^{N_r}\frac{\rho_p\lambda_{ij}^2}{\rho_p\lambda_{ij}+\sigma_w^2}$
~~~(See \eqref{eq:k_mdef_v1} and \eqref{eq:objftntracePkkMIMO_v7}.)

\STATE $\sbf_k=\sqrt{\rho_p}\ubf_{i_k}$

\STATE $\lambdabf_{i_k} \leftarrow  \sigma_w^2\lambdabf_{i_k} ./
(\rho_p\lambdabf_{i_k}+\sigma_w^2{\mathbf{1}})$  ~~~(Step *)

\ENDIF

\STATE $\lambdabf \leftarrow a^2\lambdabf +
(1-a^2)\lambdabf^{(1)}$ ~~~(Step **)

\ENDFOR \ENDWHILE
\end{algorithmic}
(Here, $./$ denotes the element-wise division and $\lambda_{ij}$
is the $j$-th element of $\lambdabf_i$. Step * incorporates the
measurement update step \eqref{eq:measurementUpdateReductionMIMO}
and Step ** incorporates the prediction step
\eqref{eq:datammseMIMO}.)
\end{algorithm}

In Algorithm \ref{alg:optimalMIMO}, the Kalman filtering error
covariance matrix $\mbox{tr}(\Pbf_{k|k})$ is minimized at each
time $k$ with the hope that such a sequence minimizes the channel
estimation MSE at the end of the pilot period of a slot. Since the
important estimation measure is the estimation error at the end of
the pilot period of each slot {(which affects the channel
estimation quality for the data transmission period under the time-varying channel assumption, as seen in \eqref{eq:timeupdateH})}, we consider
a modification to Algorithm \ref{alg:optimalMIMO} to design a
pilot beam pattern sequence, targeting at the estimation error
only at $lM+M_p$ for the $l$-th transmission block.

\begin{figure*}[!ht] {\small
\begin{align}
\text{tr}(\Pbf_{l M+M_p|l M+M_p}) &=
\sum_{ i : |\Kc_i|=1} \text{tr}(\bar{\Lambdabf}^{(l M+M_p)}_i)
+ \sum_{i : |\Kc_i|=0} \text{tr}(\bar{\Lambdabf}^{(l M+M_p)}_i) \nonumber \\
&= \sum_{i : |\Kc_i|=1}
\text{tr}\left(a^{2(lM+M_p-k^i)}\frac{\sigma_w^2\Lambdabf^{(k^i)}_i}{\rho_{k^i}\Lambdabf^{(k^i)}_i+\sigma_w^2\Ibf_{N_r}}
+ (1-a^{2(lM+M_p-k^i)})\Lambdabf^{(1)}_i  \right) + \sum_{i : |\Kc_i|=0} \text{tr}(\bar{\Lambdabf}^{(l M+M_p)}_i) \nonumber \\
& \propto \sum_{i : |\Kc_i|=1} \text{tr}\left(\frac{
a^{2(lM+M_p-k^i)}\sigma_w^2\Lambdabf^{(k^i)}_i}{\rho_{k^i}\Lambdabf^{(k^i)}_i+\sigma_w^2\Ibf_{N_r}}
\right), \label{eq:objftntracePmpBackward_power_v2}
\\~~\mbox{where } ~\Lambdabf^{(k^i)}_i&=a^{2(k^i-l
M-1)}\Lambdabf^{(lM+1)}_i+(1-a^{2(k^i-l M-1)})\Lambdabf^{(1)}_i.
\nonumber
\end{align} \vspace{-2.5em} }
\end{figure*}

\vspace{0.5em}
\begin{problem}  \label{prob:problemstatement_v2}
For each pilot symbol time $k=lM+m$ starting from 1, given
$\sbf_{i}$ for all pilot symbol time $i < k$, design $\sbf_k$ such
that
\begin{align}
\min_{\sbf_k}&~\text{tr}\left(\Pbf_{lM+M_p|k}\right)\\
\text{s.t.}&~\|\Sbf_k\|_F^2=N_r\|\mathbf{s}_k\|_2^2=N_r\rho_p,
\end{align}
where $lM+M_p$ is the end of the pilot period to which $k$
belongs.
\end{problem}
\vspace{0.5em}

Since we have
\begin{align}
\Pbf_{l M+M_p|k} &=a^{2(M_p-k)}\Pbf_{k|k} +
(1-a^{2(M_p-k)})\Rbf_{\hbf}, \label{eq:exploitDiagonalizable22}
\end{align}
the solution to Problem \ref{prob:problemstatement_v2} is given by
minimizing $\mbox{tr}(\Pbf_{k|k})$ and Algorithm
\ref{alg:optimalMIMO} can be used for this purpose too.

\subsection{Pilot Power Allocation} \label{subsec:powerallocation}

In the pilot beam pattern design in Section
\ref{subsec:ProposedMethod_1}, we considered equal pilot power for
each pilot symbol time.   We relax the equal-power constraint here
and consider the pilot beam pattern design problem again.

\begin{figure}[h]
\centerline{
\psfrag{(p1)}[c]{\footnotesize $p_1^i$} %
\psfrag{(p2)}[c]{\footnotesize $p_2^i$} %
\psfrag{(pn)}[c]{\footnotesize $p_{|\Kc_i|}^i$} %
\psfrag{(p0)}[c]{\footnotesize $\bar{p}_i$} %
\psfrag{(m1)}[c]{\footnotesize $k_1^i$} %
\psfrag{(m2)}[c]{\footnotesize $k_2^i$} %
\psfrag{(m3)}[c]{\footnotesize $k_{|\Kc_i|-1}^i$} %
\psfrag{(mn)}[c]{\footnotesize $k_{|\Kc_i|}^i$} %
\psfrag{(Mp)}[c]{\footnotesize $M_p$} %
\psfrag{(idle)}[c]{\scriptsize (a)} %
\psfrag{(pilot)}[c]{\scriptsize (b)} %
\includegraphics[scale=1.2]{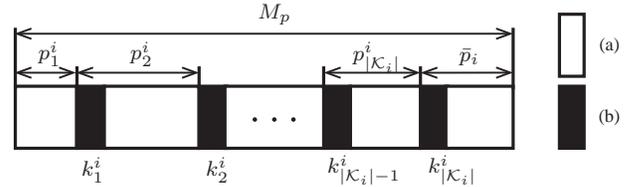}}
\vspace{-0.3em}
\caption{The use of the $i$-th transmit eigenvector $\ubf_i$ as
the pilot beam in a slot where $k_j^i\in \Kc_i$ and $\bar{p}_i
=M_p-\|\pbf_i\|_1$: (a) $\ubf_i$ is not used  and (b) $\ubf_i$ is
used.} \label{fig:useOfnthPilotBeam}\vspace{-0.5em}
\end{figure}

First, we will derive  a necessary condition of an optimal pilot
beam sequence that is useful for further pilot design. (This
condition is given in Proposition \ref{prop:powerallocation}.) To
do so, let us first define some notations.  For $1\le i \le N_t$,
let $\Kc_i =\{k|  \sbf_k=\sqrt{\rho_k}\ubf_i
\}=\{k_1^i,k_2^i,\cdots, k_{|\Kc_i|}^i\}\subset\{l M+1,\ldots, l
M+M_p\}$ be the time index set in the $l$-th slot for which the
$i$-th transmit eigenvector $\ubf_i$ (obtained from
$\Rbf_t=\Ubf\Sigmabf\Ubf^H$ and $\Ubf=[\ubf_1,\cdots,\ubf_{N_t}]$)
is used as the pilot beam pattern. Note that some eigenvectors may
not be used as the pilot beam pattern depending on the channel
statistics. Under the assumption that the transmitter has  total
power $M_p\rho_p$ for the pilot transmission period, we denote by
$\rho_{k_j^i}$  the pilot signal power for the use of the $i$-th
transmit eigenvector at time $k_j^i\in\Kc_i$ and define a pilot
interval vector $\pbf_i =\bigl[p_1^i,p_2^i,\cdots,p_{|\Kc_i|}^i\bigr]^T$ as
shown in Fig. \ref{fig:useOfnthPilotBeam}. The following
proposition provides a property regarding optimal pilot power
allocation.

\vspace{0.5em}
\begin{proposition}\label{prop:powerallocation}
An optimal pilot beam pattern sequence minimizing
$\text{tr}(\Pbf_{l M+M_p|l M+M_p})$ in the $l$-th slot should
satisfy the condition that all the pilot power for a transmit
eigen-direction is allocated to the last use of the
eigen-direction  in the slot. That is, one transmit
eigen-direction should not appear more than once in the pilot
period of each slot.
\end{proposition}
\vspace{0.5em}
{\em Proof:} See Appendix \ref{app:powerallocation}.

\vspace{0.5em}

Now consider the problem of  joint design of beam patten index
selection and power allocation.  As seen in Section
\ref{subsec:ProposedMethod_1}, the pilot beam pattern sequence
design is a difficult problem even with fixed pilot power. In the
case of pilot beam pattern sequence design with power control, we
have a more complicated situation. Our approach to this
complicated joint design problem is to separate the beam pattern
index selection and the power allocation, although it is
suboptimal.  We again use the sequential beam pattern index
selection based on \eqref{eq:k_mdef_v1} together with Proposition
\ref{prop:powerallocation}, but now we do not know the allocated
pilot power  beforehand. To circumvent this difficulty, we exploit
the property of the argument in \eqref{eq:k_mdef_v1}. Note that
the argument in \eqref{eq:k_mdef_v1} is an increasing\footnote{
 A real-valued function $\phi$
defined on some set $\Hc$ of $n\times n$ Hermitian matrices is
{\it increasing} on $\Hc$ if $\Abf  \preceq \Bbf  \Rightarrow
\phi(\Abf)\le  \phi(\Bbf)$,  whenever $\Abf,\Bbf\in\Hc$ \cite[Ch.
16]{Marshall&Olkin:book}. } function of $\Lambdabf^{(k)}_i$ for
any positive $\rho_p$. Hence, if we choose
$\Lambdabf_{i^\prime}^{(k)}$ s.t. $\Lambdabf_{i^\prime}^{(k)}
\succeq \Lambdabf_{i}^{(k)}$ for all $i\ne i^\prime$, this index
$i^\prime$ is optimal. Note that for this selection method, we do
not need the knowledge of the current pilot power $\rho_k$ at time
$k$ ($\rho_p$ in the case of \eqref{eq:k_mdef_v1}). However, there
may not be such an index and hence, we replace this majorization
criterion with a simple trace criterion since all the elements
$\Lambdabf^{(k)}_i$ are non-negative. (Having the maximum trace is
at least a necessary condition for being the majorizing index.)
Based on this, we propose to choose the beam pattern index at time
$k$ to minimize $\mbox{tr}(\Pbf_{k|k})$ (or equivalently
$\mbox{tr}(\Pbf_{lM+M_p|k})$ as follows. First, consider time
$k=lM+1$ under the assumption that the pilot sequence and power is
already determined for the previous slots. We choose $ i_1 :=
\argmax_{i} ~\text{tr}(\Lambdabf^{(1)}_i)$. With the first index
selected, consider $k=lM+2$. Now, applying the condition of
Proposition \ref{prop:powerallocation}, we choose $i_2 :=
\argmax_{i \notin \{i_1\}} ~\text{tr}(\Lambdabf^{(2)}_i)$. This is
possible without knowing $\rho_{lM+1}$ since only
$\Lambdabf^{(2)}_{i_1}$ is affected by $\rho_{lM+1}$ and $i_1$ is
not considered from $k \ge lM+2$. Then, we proceed to $k=lM+3$. In
this way, we can choose $i_1,\cdots,i_{M_p}$ without knowing
$\rho_{lM+1},\cdots,\rho_{lM + M_p}$ based on the trace criterion
and Proposition \ref{prop:powerallocation}. For a selected index
$i$, $\Kc_i=\{k_1^i\}$ and for an unselected index $i$, $\Kc_i =
\emptyset$. Then, we have $\sum_{i=1}^{N_t} |\Kc_i|\le M_p$. (Let
us use $k^i$ for $k_1^i$.) Once $i_1,\cdots,i_{M_p}$ are
determined, the optimization goal
$\mbox{tr}(\Pbf_{lM+M_p|lM+M_p})$ is given by
\eqref{eq:objftntracePmpBackward_power_v2}.

Based on \eqref{eq:objftntracePmpBackward_power_v2}, the pilot
power  optimization problem is formulated as
\begin{align}
\min_{\boldsymbol\rho  }& \sum_{i:|\Kc_i|=1} \text{tr}\left(\frac{ a^{2(l M+M_p-k^i)} \sigma_w^2 \Lambdabf^{(k^i)}_i }{\rho_{k^i}\Lambdabf^{(k^i)}_i + \sigma_w^2\Ibf_{N_r}}  \right)  \label{eq:optprobPowerAllocation} \\
\text{s.t.}&~ \|\rhobf\|_1  = M_p \rho_p,~ \rho_{k^i}\ge 0,
\label{eq:objftntracePmpBackward_power_Constraint}
\end{align}
where $\rhobf=[\rho_{lM+1},\ldots,\rho_{lM+M_p}]^T$. The problem
\eqref{eq:optprobPowerAllocation} can be solved by  water-filling
power allocation\cite{Cover&Thomas:book} (see Appendix
\ref{subsec:poweralloation} for details), and the corresponding
algorithm is summarized in Algorithm \ref{alg:optimalMIMObackwardPower}.
In the MIMO case, $\rhobf$
needs to be solved numerically from \eqref{eq:OptSolutionNu},
whereas in the MISO case we have a closed-form solution given by
\begin{align}
\rho_{k^i} = \left(a^{l M+M_p-k^i}\frac{\sigma_w}{\sqrt{\nu}} -
\frac{\sigma_w^2}{\lambda^{(k^i)}_i}\right)^+,
\end{align}
where $(\cdot)^+ = \max(\cdot, 0)$ and $\nu$ is evaluated from the
power constraint
\eqref{eq:objftntracePmpBackward_power_Constraint}.

In high and low SNR regimes, the optimal power allocation  can be
approximated by simpler forms:

\vspace{0.2em}
\noindent {\it Case 1) High SNR:
$\rho_{k^i}\lambda^{(k^i)}_{{ij}}\gg \sigma_w^2$}
\begin{align}   \label{eq:highSNRsolution}
\rho_{k^i} = \frac{M_p\rho_p(1-a)}{1-a^{M_p}}a^{M_p-k^i},
\end{align}
where $\lambda^{(k^i)}_{{ij}}$ is the $j$-th diagonal element of
$\Lambdabf^{(k^i)}_i$. \vspace{0.2em}

\noindent {\it Case 2) Low SNR:
$\rho_{k^i}\lambda^{(k^i)}_{{ij}}\ll \sigma_w^2$}
\begin{align}
\rho_{k^{i^\prime}} &= M_p\rho_p  \label{eq:lowSNRsolution}\\
i^\prime &= \textstyle\argmax_{i:|\Kc_i|=1} ~
\text{tr}\left(a^{2(lM+M_p-k^i)} \Lambdabf^{(k^i)}_i  \right).
\nonumber
\end{align}

In the special case of static channels, i.e., $a=1$, the proposed
power allocation strategy covers the result of Kotecha and
Sayeed\cite{Kotecha&Sayeed:04SP}, which considers the MMSE channel
estimation with power control for quasi-static channels.

\subsection{Block-fading Channel Model} \label{subsec:blockading}

In this subsection, we consider a block Gauss-Markov fading
channel model under which the channel is constant for each  slot,
i.e., $\hbf_k=\hbf_{l}$ for $k=l M + m~ ( m=1,2,\cdots,M)$, but
varies continuously across slots according to $\hbf_{l+1} =
a\hbf_{l} + \sqrt{1-a^2}\bbf_{l}$. We assume that the base station
equipped with $N_t$ antennas serves a single-antenna terminal for
simplicity\cite{Marzetta:10WCOM};   each coherence time block of
$M$ symbols is composed of a training period of $M_p$  symbols and
a data transmission period of $M_d$ symbols; and $M_p < N_t$. By
stacking $M_p$ symbols during the $l$-th training period, we have
the received signal $\ybf_{l}\in\mathbb{C}^{M_p}$, given by
\begin{align}
\ybf_{l} &= \Sbf_{l}^H\hbf_{l} + \wbf_{l},
\end{align}
\begin{algorithm} [t]                       
\caption{Sequential Pilot Beam Pattern Design with Power Allocation}          
\label{alg:optimalMIMObackwardPower}  
\begin{algorithmic}                          
\REQUIRE Perform the ED of $\Rbf_t=\Ubf\Sigmabf\Ubf^H$ and
$\Rbf_r=\Vbf\Gammabf\Vbf^H$ where $\Rbf_\hbf=\Rbf_t\otimes\Rbf_r$.
Store $\lambdabf^{(1)}=\text{diag}(\Sigmabf \otimes \Gammabf)$,
and $\Ubf=[\ubf_1,\cdots,\ubf_{N_t}]$.

\STATE $\lambdabf=\lambdabf^{(1)}$ and partition
$\lambdabf=[\lambdabf_1^T,\cdots,\lambdabf_{N_t}^T]^T$

\WHILE{$l =0,1,\cdots$}

\STATE $\Kc_i=\emptyset$ for $1 \le i\le N_t$

\FOR{$m = 1$ to $M_p$}

\STATE $k=lM+m$

\STATE $i_k=\argmax_{i: \mbox{{\footnotesize not used in this
slot}}} \sum_{j} \lambda_{ij}$, where $\lambda_{ij}$ is the $j$-th
element of $\lambdabf_i$, i.e., $\sum_j
\lambda_{ij}=\mbox{tr}(\Lambdabf_i)$.

\STATE Set $\Kc_{i_k}=k$

\STATE $\lambdabf \leftarrow a^2\lambdabf +
(1-a^2)\lambdabf^{(1)}$

\ENDFOR

\STATE Obtain the power allocation $\rhobf$ by solving
\eqref{eq:optprobPowerAllocation}.

\FOR{$i = 1$ to $N_t$}

\IF{$|\Kc_i|=1$}

\STATE $\sbf_{k^i} = \sqrt{\rho_{k^i}}\ubf_{i_k}$

\ENDIF

\ENDFOR

\FOR{$m = 1$ to $M$}

\STATE Perform Kalman measurement update and prediction with the
obtained $\{\sbf_k\}$ to track the correct error covariance
matrix.

\ENDFOR

\ENDWHILE

Note that in the first for-loop, the measurement update step is
not implemented since we do not choose the used eigen-direction
index again and thus we only need the prediction steps to select
the eigen-direction indices.
\end{algorithmic}
\end{algorithm}\noindent
where $\ybf_{l}=[y_{l M+1},\ldots,y_{l M+M_p}]^T$ and
$\Sbf_{l}=[\sbf_{l M +1} \cdots \sbf_{l M +M_p}]$. We further
assume that $\Sbf_{l}^H\Sbf_{l} = \rho_p\Ibf_{M_p}$
\cite{Hassibi&Hochwald:03IT,Santipach&Honig:10IT}.
The following proposition provides a property of  optimal $\Sbf_{l}$ under the
block-fading channel model.

\vspace{0.5em}
\begin{proposition}\label{pro:argminMMSE_blkfading}
Given all previous pilot signals $\Sbf_{l^\prime}$ ($l^\prime <
l$), the pilot beam signal $\Sbf_l$ at the $l$-th training period
minimizing $\text{tr}(\Pbf_{l | l})$ is given by the scaled version of the $M_p$
dominant eigenvectors of the Kalman prediction error covariance matrix
$\Pbf_{l|l-1}$ for the $l$-th
training period.
\end{proposition}
\vspace{0.5em}
{\em Proof:} See Appendix \ref{app:argminMMSE_blkfading}.

\vspace{0.5em}

As in the symbolwise Gauss-Markov channel model, all Kalman
prediction error covariance matrices that are used for the
orthogonal pilot beam pattern design have the same set of
eigenvectors of $\Rbf_{\hbf}$, i.e., $\Rbf_{\hbf}$, $\Pbf_{l | l}$
and $\Pbf_{l|l^\prime}$ are {\em simultaneously diagonalizable}.
(Proof is omitted since it can be shown similarly as in
Proposition \ref{pro:periodicityPilotPattern_mimo}.) Thus, the
proposed algorithm in the previous section can easily be extended
to the block-fading Gauss-Markov channel model. Previously, it was
proposed by some other researchers that the $M_p$ dominant
eigenvectors of $\Rbf_\hbf$ are used for the $M_p$ pilot symbol
times for every slot under the block i.i.d. fading model
\cite{Kotecha&Sayeed:04SP}. However, in our proposed method, we
use for the $M_p$ pilot beam patterns in the $l$-th slot the $M_p$
dominant eigenvectors of $\Pbf_{l|l^\prime}$ instead of
$\Rbf_\hbf$ to incorporate channel dynamics and to track the most
efficient $M_p$ eigen-directions over time. Note that the full set
of eigenvectors is the same for $\Rbf_\hbf$ and
$\Pbf_{l|l^\prime}$ and that $\Rbf_\hbf$ does not change over time
under the considered {\em stationary} Gauss-Markov channel model.
This tracking feature of the proposed method yields a significant
gain over the previous method in time-varying channels when the
channel dynamic is known, as seen in Section \ref{sec:numericalresult}.

\vspace{-0.3em}
\section{{Discussion: Practical Implementation and Multi-User Scenario}} \label{sec:discussion}

{In this section, we make some comments relative to practical implementation of our proposed pilot design and channel estimation scheme in real-world massive MIMO systems.}

{First, consider the type and amount of feedback necessary for a massive MIMO system. 
One approach is to have the mobile station estimate the full channel state vector and feed that
back to the base station.  For a massive MIMO system, this approach requires a large amount of feedback and may be difficult to implement in practice.  Alternatively,  the mobile station may simply feed back the received signal $\ybf_k\in\mathbb{C}^{N_r}$ at each time instant, i.e., have the mobile station effectively transmit back the inner product between the current beamforming vector and the current channel state vector plus noise, and use that information to form an estimate of the channel at the base station \cite{Noh&Zoltowski&Sung&Love:13ASILOMAR}.  The latter method is more effective in terms of the amount of feedback and does not require any modifications to the algorithm proposed in this paper.
}

{Second,  consider the estimation of the channel fading coefficient $a$ in the channel time-varying model (\ref{eq:statespacemodel_h}). Since $a$ depends on the mobile speed of the receiver, it can be estimated by using the uplink received signal directly \cite{Iltis:90COM,Tsatsanis&Giannakis&Zhou:96ICASSP,Dai&Zhang&Xu&Mitchell&Yang:12EURASIP,Liu&Hansson&Vandenberghe:13SCL}. (A simple correction due to the uplink and downlink carrier frequency difference in FDD systems should be applied.) This problem falls into the general area of system identification of state-space models. Especially, blind techniques based on subspace approaches can be applied here. Interested readers are referred to \cite[Section 2]{Liu&Hansson&Vandenberghe:13SCL}.}


{Next, throughout the paper, we assume that the downlink channel covariance matrix $\Rbf_{\hbf}$ is known to the system.  If $\Rbf_\hbf$ is estimated at the receiver (mobile station) and fed back to the base station through some control channel, the feedback overhead may be significant. Fortunately, there exist methods that can circumvent this difficulty.
One way is to estimate the downlink channel covariance matrix $\Rbf_{\hbf}$ from the uplink channel covariance matrix \cite{Raleigh&Diggavi&Jones&Paulraj:95ICC,Liang&Chin:01JSAC,Hochwald&Marzetta:01SP}.\footnote{Note that in the MISO downlink case, the uplink is SIMO. In the time-domain duplex (TDD) case, the uplink and downlink channel covariance matrices are the same.}
The downlink $\Rbf_{\hbf}$ can be estimated from the uplink channel covariance matrix even though they are a bit separated in the frequency domain in the FDD case. Interested readers are referred to \cite{Raleigh&Diggavi&Jones&Paulraj:95ICC,Liang&Chin:01JSAC,Hochwald&Marzetta:01SP}.}

{Furthermore, we here propose even a simpler method to obtain $\Rbf_\hbf$ based on the one ring model and the Toeplitz distribution theorem for 1-dimensional or 2-dimenional large uniform arrays.
 Consider a 1-dimensional large uniform array with $N_t$ antenna elements for simplicity. Each element of the array performs spatial-sampling of the signal. Thus, if we view these spatial samples as discrete-time samples, the conventional (discrete-time) frequency domain  corresponds to the virtual angle domain.\footnote{The virtual angle $\xi$ is related to the physical angle $\theta$ by $\xi = \frac{d}{\lambda}\sin(\theta)$, where $d$ is the antenna spacing and $\lambda$ is the carrier wavelength. When $d/\lambda=1/2$, $-\frac{\pi}{2} \le \theta \le \frac{\pi}{2}$ corresponds to $-\frac{1}{2} \le \xi \le \frac{1}{2}$.}
For the one-ring model with a uniform array under a far-field assumption, the channel covariance matrix $\Rbf_{\hbf}$ is Toeplitz\cite{Adhikary&Nam&Ahn&Caire:13IT}.  It is known that when the size of a Toeplitz covariance matrix is large,  the Toeplitz matrix can be eigen-decomposed by a DFT matrix, which is known as the Toeplitz distribution theorem \cite{Grenander&Szego:book,Sungetal:09IT,Adhikary&Nam&Ahn&Caire:13IT}, i.e., $\Rbf_{\hbf} \approx \Fbf \Dbf \Fbf^H$ 
where $\Fbf$ is a DFT matrix and $\Dbf$ is a diagonal matrix that contains the virtual angular power spectral values. (This is why the eigen-decomposition of a Toeplitz covariance matrix is also called the spectral decomposition.) For a one-ring model with angle-of-arrival (AoA) and angle-dispersion ($\Delta$), the elements of $\Dbf$ are non-zero only for the angle spectrum $(\mbox{AoA}-\Delta,~\mbox{AoA}+\Delta)$. Thus, when AoA and $\Delta$ are given, $\Rbf_\hbf$ can be constructed from the corresponding columns of $\Fbf$ and the angular power spectral values.  Note that the $k$-th column of $\Fbf$ is given by
\begin{equation} \label{eq:onebeamdirection}
\frac{1}{\sqrt{N}}[1, e^{\iota 1 \xi_k 2\pi/N},  e^{\iota 2 \xi_k 2\pi/N}, \cdots,e^{\iota (N-1) \xi_k 2\pi/N}]^H.
\end{equation}
This is  simply the steering vector for the physical angle $\theta_k = \sin^{-1}(\xi_k\lambda/d)$.  Under the model, the channel is given by a random linear combination of column vectors or steering vectors with the form (\ref{eq:onebeamdirection}) looking at the angle range $(\mbox{AoA}-\Delta,~\mbox{AoA}+\Delta)$. (Channel estimation in the previous sections is nothing but estimation of  these random linear combination coefficients.) 
 The AoA can be estimated from the uplink signal model (there are numerous practical AoA or DoA estimation algorithms) and $\Delta$ can be pre-measured or predetermined for each carrier frequency by reflecting the typical scattering environment. The angular power spectrum  can also be estimated based on one of typical spectral estimation methods \cite{Brockwell&Davis:book}. Here, the angular power spectrum is estimated by using the uplink signal and a correction similar to those in \cite{Raleigh&Diggavi&Jones&Paulraj:95ICC,Liang&Chin:01JSAC,Hochwald&Marzetta:01SP} can be applied to obtain a downlink counterpart.
Simulations will be presented towards the end of the next section in which the pilot beam patterns are approximated by DFT vectors without much loss in performance.
}
 
{In summary, the proposed pilot design and channel estimation method can be run in the following practical way:
  \begin{enumerate}
     \item first estimate the AoA based on the uplink signal and selects the columns of $\Fbf$ corresponding to $(\mbox{AoA}-\Delta,~\mbox{AoA}+\Delta)$; %
     \item estimate the angular power profile for $(\mbox{AoA}-\Delta,~\mbox{AoA}+\Delta)$ from the uplink channel response \cite{Brockwell&Davis:book}, and finally obtain a downlink power profile via correction \cite{Raleigh&Diggavi&Jones&Paulraj:95ICC,Liang&Chin:01JSAC,Hochwald&Marzetta:01SP}. This downlink angular power profile gives $\lambdabf^{(1)}$ in Algorithm 1; %
     \item   estimate the mobile speed of the terminal (i.e., $a$) based on the uplink by using one of system identification algorithms \cite{Iltis:90COM,Tsatsanis&Giannakis&Zhou:96ICASSP,Dai&Zhang&Xu&Mitchell&Yang:12EURASIP,Liu&Hansson&Vandenberghe:13SCL}; and %
     \item finally run one of the algorithms in the previous sections. (By reciprocity, the AoA and the terminal velocity are the same for the up and down links.)%
 \end{enumerate}}

{Finally, consider the multi-user case. Note that the system model \eqref{eq:statespacemodel_y1} is for a single-user MIMO channel. However,  many of current real-world wireless communication systems as those in 3GPP support user-dedicated pilot and control channels in addition to a common pilot and control channel for effective channel estimation for each user. Thus, the proposed method can be applied to these dedicated pilot channels. Furthermore, the proposed method can well be combined with the recently proposed joint spatial division and multiplexing (JSDM) framework for multiuser massive MIMO systems \cite{Adhikary&Nam&Ahn&Caire:13IT}.
In the JSDM,
the multiple users (MU) in a sector are partitioned into groups each of which has  approximately the same channel covariance matrix. (Each set of the partition can be viewed as a virtual subsector.)
Here, if the groups or subsectors are sufficiently well separated in the AoA domain, the dominant eigenvectors of the channel covariance matrices become linearly independent for different groups.
To serve MU-MIMO in the same time-frequency slot, we can choose the users that have non-overlapping supports of their AoA distribution as  in \cite{Adhikary&Nam&Ahn&Caire:13IT}.
Then, the optimal pilot beam patterns become different and orthogonal among non-overlapping groups.
In this case, the system model \eqref{eq:statespacemodel_y1} can be regarded as the signal model for a scheduled user in one of the non-overlapping subsectors of the overall multi-user downlink.}

\vspace{-0.3em}
\section{Numerical Results} \label{sec:numericalresult}

In this section, we provide some numerical results to evaluate the
performance of the proposed algorithms.  We considered $N_t\in\{32,250\}$  transmit antennas and $N_r\in\{1,2\}$ receive antennas for our
massive MIMO systems. We adopted $2.5GHz$ carrier frequency and
$100\mu s$ symbol duration  with a typical mobile speed range
from $v=3km/h$ $(a=0.9999)$  to $30km/h$ $(a=0.9995)$. For all
considered pilot design methods, we used Kalman filtering and
prediction for the channel estimator. To evaluate the channel estimation
performance, we computed the normalized mean square error (NMSE),
given by $\frac{1}{\text{tr}(\Rbf_{\hbf})}\text{tr}(\Pbf_{k|k})$.
The pilot symbol SNR was defined as $\rho_p/\sigma_w^2$, the data
symbol SNR was defined as $\rho_d/\sigma_w^2$, and the two SNR
values were the same throughout the simulation. The noise variance
$\sigma_w^2$ was determined according to the SNR value with
$\rho_p=\rho_d=1$, and the received SNR is defined as \eqref{eq:receivedSNRrevision},
which incorporates the effect of beamforming gain and imperfect
channel estimation. The channel estimation performance for each of
the considered methods was averaged over $1,000$ Monte Carlo runs.

\begin{figure}[!t]
\vspace{-0.55em}
\centerline{\includegraphics[scale=0.52]{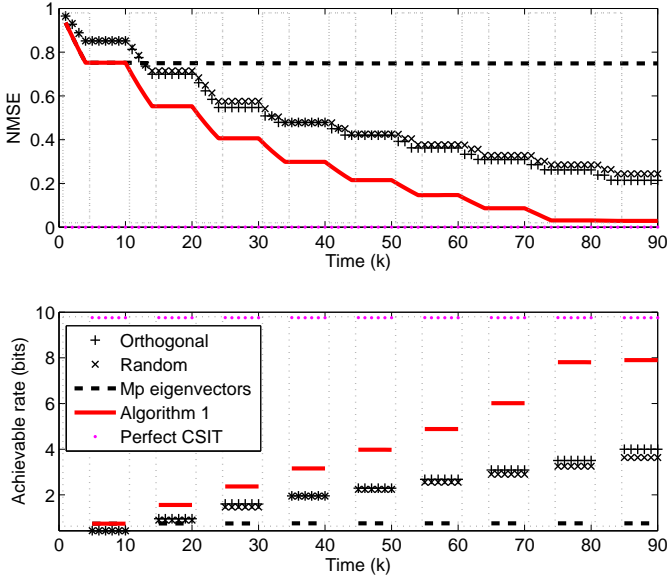}}
\vspace{-1.7em} \caption{NMSE and a lower bound on achievable rate versus time
index $k$:  $M=10$,  $M_p=4$, $\sigma_w^2=10^{-1.5}$, $N_t=32$, $N_r=2$, $r=0.6$, and $v=3km/h$ (The dotted rectangles denote pilot transmission periods.)}
\label{fig:nmserate_expo_snr15db_3kmh} \vspace{-1.6em}
\end{figure}

{First, we considered the {\em exponential correlation model} for
channel spatial correlation, given by $[\Rbf_t]_{i,j}=
r_t^{2|i-j|}$ and $[\Rbf_r]_{i,j}=
r_r^{2|i-j|}$, where $r_t$ and $r_r$ are the transmit and receive correlation coefficients between two
adjacent antenna elements, respectively ($r_t=r_r=r$ for simplicity).
Since the phase of $r$ is irrelevant to the eigenvalues of $\Rbf_\hbf$, we assume without loss of generality  that the phase of $r$ is fixed to be zero (i.e., $r\in\mathbb{R}$)}.
Fig. \ref{fig:nmserate_expo_snr15db_3kmh} shows the channel
estimation performance of several pilot pattern design methods
\cite{Santipach&Honig:10IT} for the exponential channel correlation model with $r=0.6$, $N_t=32$, and $N_r=2$.  
The performance of the $M_p$ dominant eigenvectors of $\Rbf_{\hbf}$ as the $M_p$ pilot beam
patterns for every pilot period is also shown. 
{
It is seen that the proposed algorithm tracks the channel state fast due to the ability of the proposed
method's tracking the spectral distribution of the channel MSE.
Thus, the proposed method converges more quickly. 
The use of orthogonal or random beam patterns (which span the overall space) yields
reasonable performance with slightly increased convergence time
compared to the proposed method. In the case of the fixed
$M_p$ dominant eigenvectors of $\Rbf_\hbf$ for the pilot beam
pattern in every pilot period, one can only minimize the channel
MSE along the fixed $M_p$ eigen-directions, and the coverage of
only $M_p$ fixed eigen-directions in the space is not enough for
very large $N_t$ when $M_p$ is small. Hence, the channel
estimation MSE performance of the fixed pilot beam pattern method
is saturated quickly. 
By replacing the channel estimation error plus noise with independent additive Gaussian noise during the data transmission phase \cite{Hassibi&Hochwald:03IT}, we showed the training-based lower bound on achievable data rate in Fig. \ref{fig:nmserate_expo_snr15db_3kmh}.
The proposed method also guarantees a good (average) lower bound on achievable rate due to precise channel estimation.
}

\begin{figure}[!t]
\centerline{\includegraphics[scale=0.42]{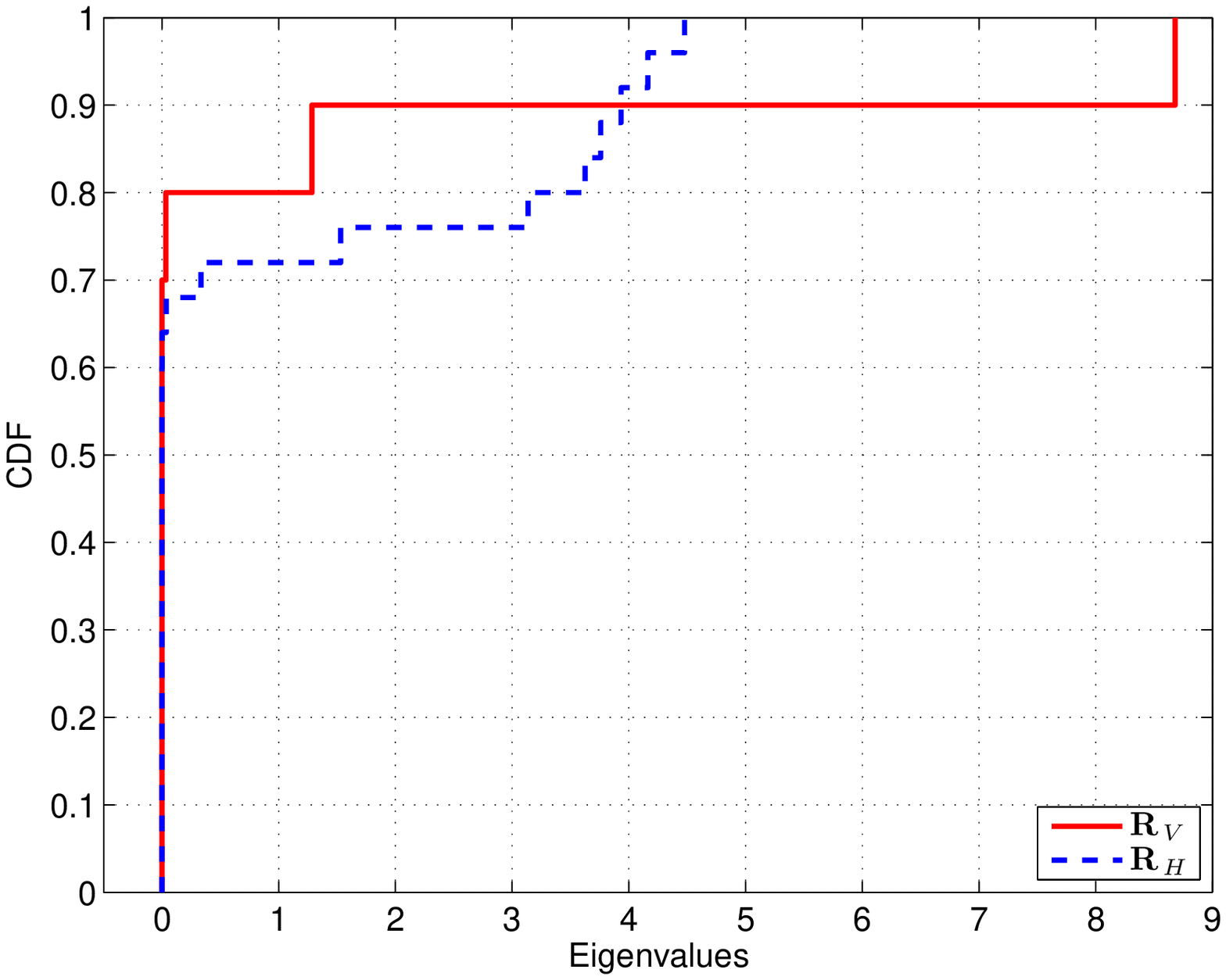}}
\vspace{-1.0em} \caption{Empirical eigenvalue CDF of $\Rbf_V$ and $\Rbf_H$} \label{fig:eigenValuesPlanar} \vspace{-1.12em}
\end{figure}

Next, we considered the (more realistic) {\em one-ring} channel
model which well models typical cellular configurations
\cite{Shiu&Foschini&Gans&Kahn:00COM,Adhikary&Nam&Ahn&Caire:13IT}.
The channel spatial correlation with a ULA is given by
\eqref{eq:channelCovOneRing} and depends on AoA $\theta$ and AS
$\Delta$, and this model can be extended to the 2-dimensional
array case (See \cite{Shiu&Foschini&Gans&Kahn:00COM} for details.)
Indeed, we considered a transmitter employing a $10\times 25$
uniform planar array (UPA) on half-wavelength lattice,
$D=\frac{1}{2}$ with $N_r=1$. In order to compute the vertical and horizontal
channel covariance matrices $\Rbf_{V},\Rbf_{H}$, we assume that
the transmit antenna is located at an elevation of $h=60m$, the
scattering ring of the receiver has radius $r=30m$, and the
distance from the transmitter is $s=100m$. The path loss between
the transmitter and the receiver is given by
$(1+(\frac{s}{d_0})^\alpha)^{-1}$, where the path loss exponent is
set as  $\alpha=3.8$ and the reference distance is set as
$d_0=30m$. Then, the parameters for the channel covariance
matrices  $\Rbf_{V}$ and $\Rbf_{H}$ are given by
$\Delta_V=\frac{1}{2}\left(\arctan(\frac{s+r}{h})-\arctan(\frac{s-r}{h})\right)$,
$\theta_V=\frac{1}{2}\left(\arctan(\frac{s+r}{h})+\arctan(\frac{s-r}{h})\right)$,
$\Delta_H=\arctan(\frac{r}{s})$, and $\theta_H=\frac{\pi}{6}$.
Finally, the channel covariance matrix is given by $\Rbf_{\hbf} =
\Rbf_{H}\otimes\Rbf_{V}$ \cite{Adhikary&Nam&Ahn&Caire:13IT}.
Fig. \ref{fig:eigenValuesPlanar}  shows the empirical cumulative
distribution function (CDF) of the eigenvalues of $\Rbf_{\hbf}$
obtained in the above, and  exhibits rank-deficiency in the
spatial channel covariance matrices due to local scattering around
the receiver. Note that 70 \% to 80\% of the eigenvalues are zero.

\begin{figure}[!t]
\centerline{\includegraphics[scale=0.42]{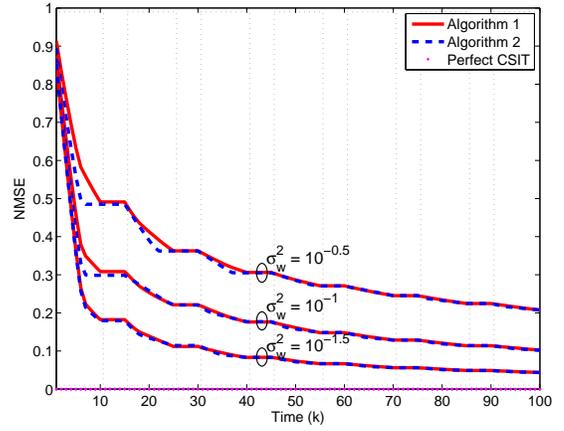}}
\vspace{-1.0em} \caption{NMSE versus time index $k$ where $M=15$, $M_p=10$, and $v=3km/h$}
\label{fig:nmsesnr_planar_snr51015db_3kmh} \vspace{-1.7em}
\end{figure}
\begin{figure*}[!t]
\centerline{ \SetLabels 
\L(0.19*-0.03) \footnotesize (a) Transient tracking \\
\L(0.68*-0.03) \footnotesize (b) Steady-state tracking \\
\endSetLabels
\leavevmode
\strut\AffixLabels{
\includegraphics[scale=0.46]{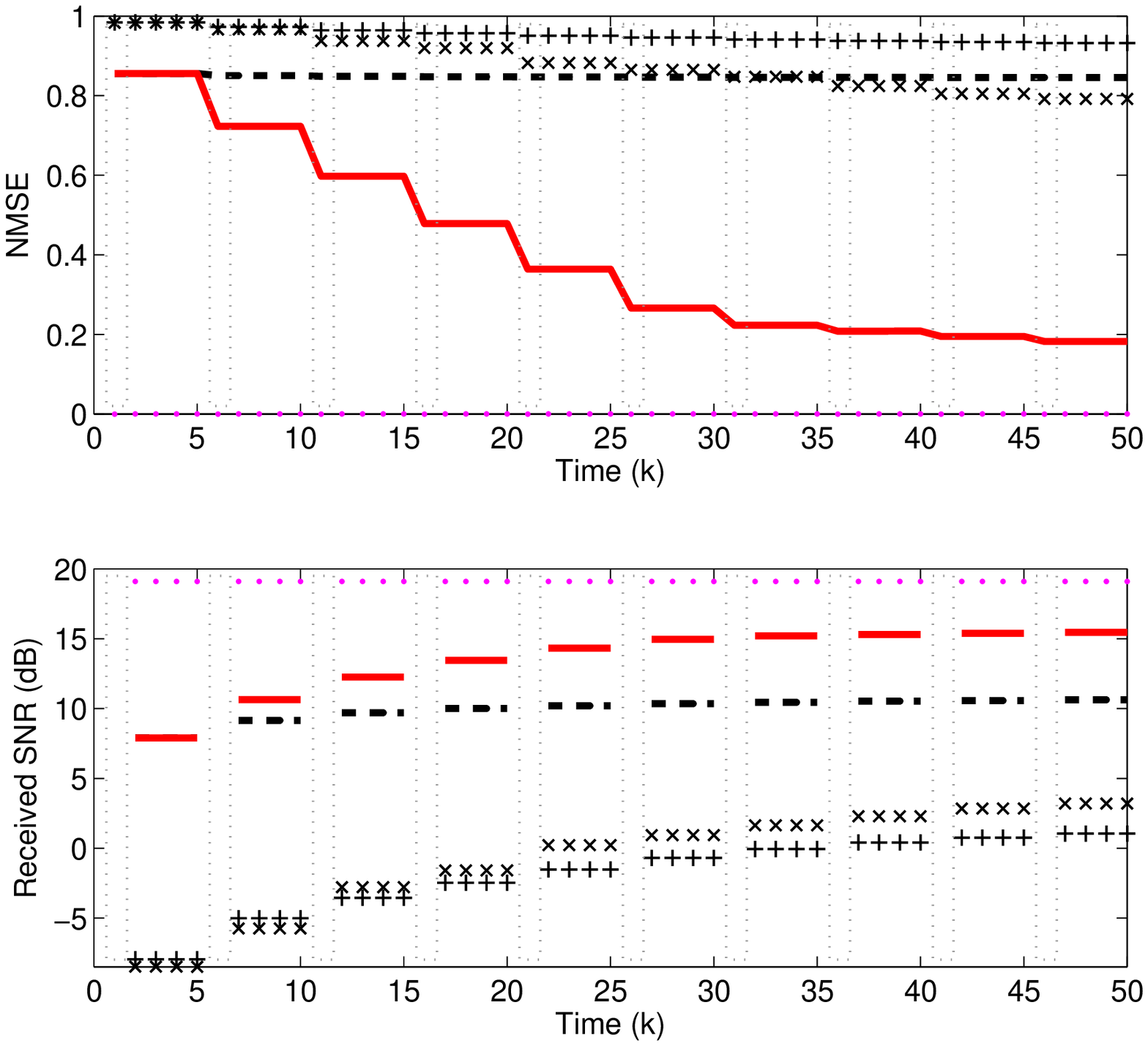}\hspace{1em}
\includegraphics[scale=0.46]{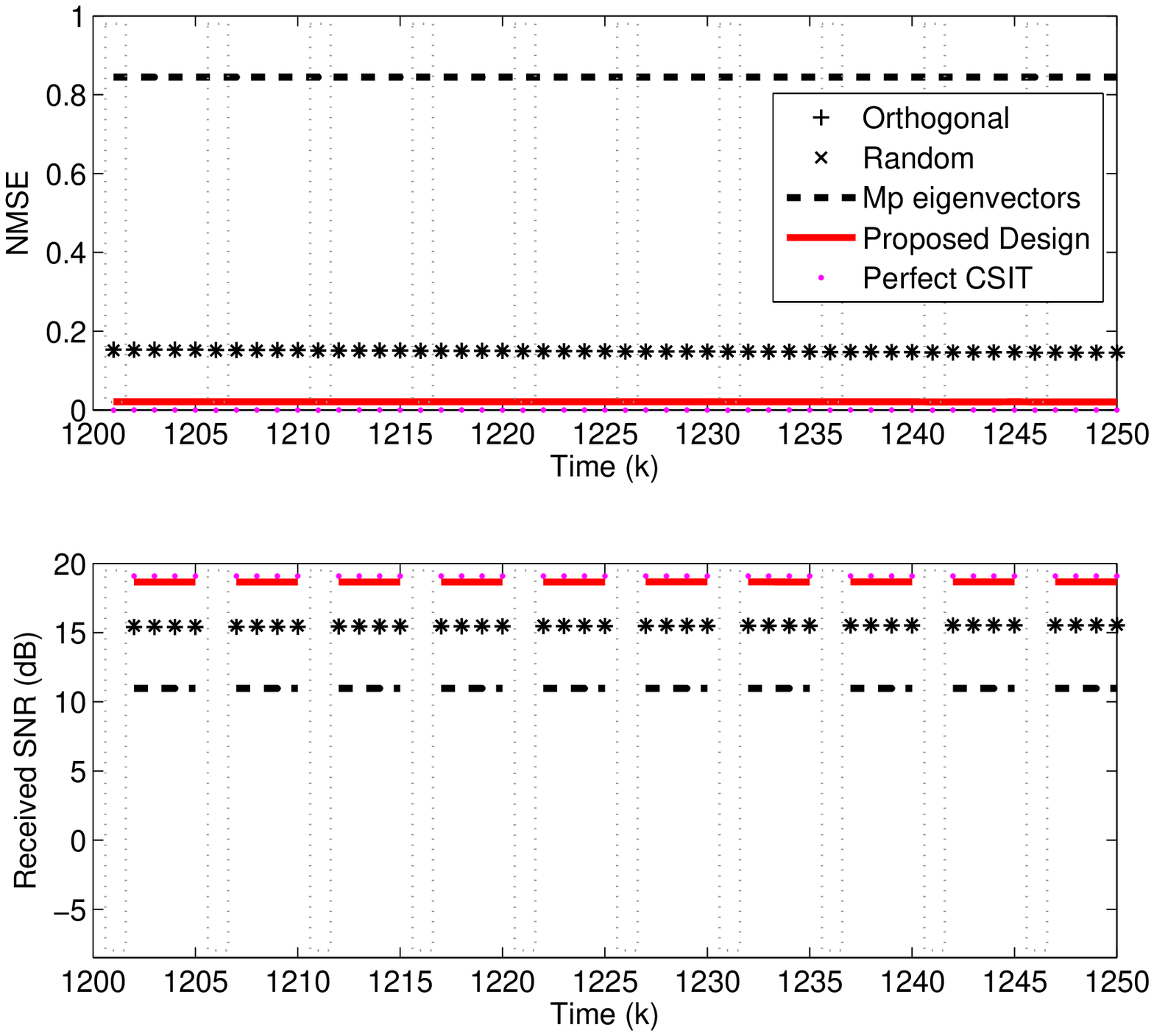}}}
\vspace{-0.1em} \caption{NMSE and received SNR versus time index $k$ where $M=5$, $M_p=1$, $\sigma_w^2=10^{-1.5}$, and $v=3km/h$}
\label{fig:nmsesnr_planar_snr15db_3kmh} \vspace{-1.2em}
\end{figure*}
\begin{figure}[!t]
\centerline{ \SetLabels
\L(0.36*-0.05) \footnotesize (a) QPSK modulation \\
\endSetLabels
\leavevmode \strut\AffixLabels{
\includegraphics[scale=0.40]{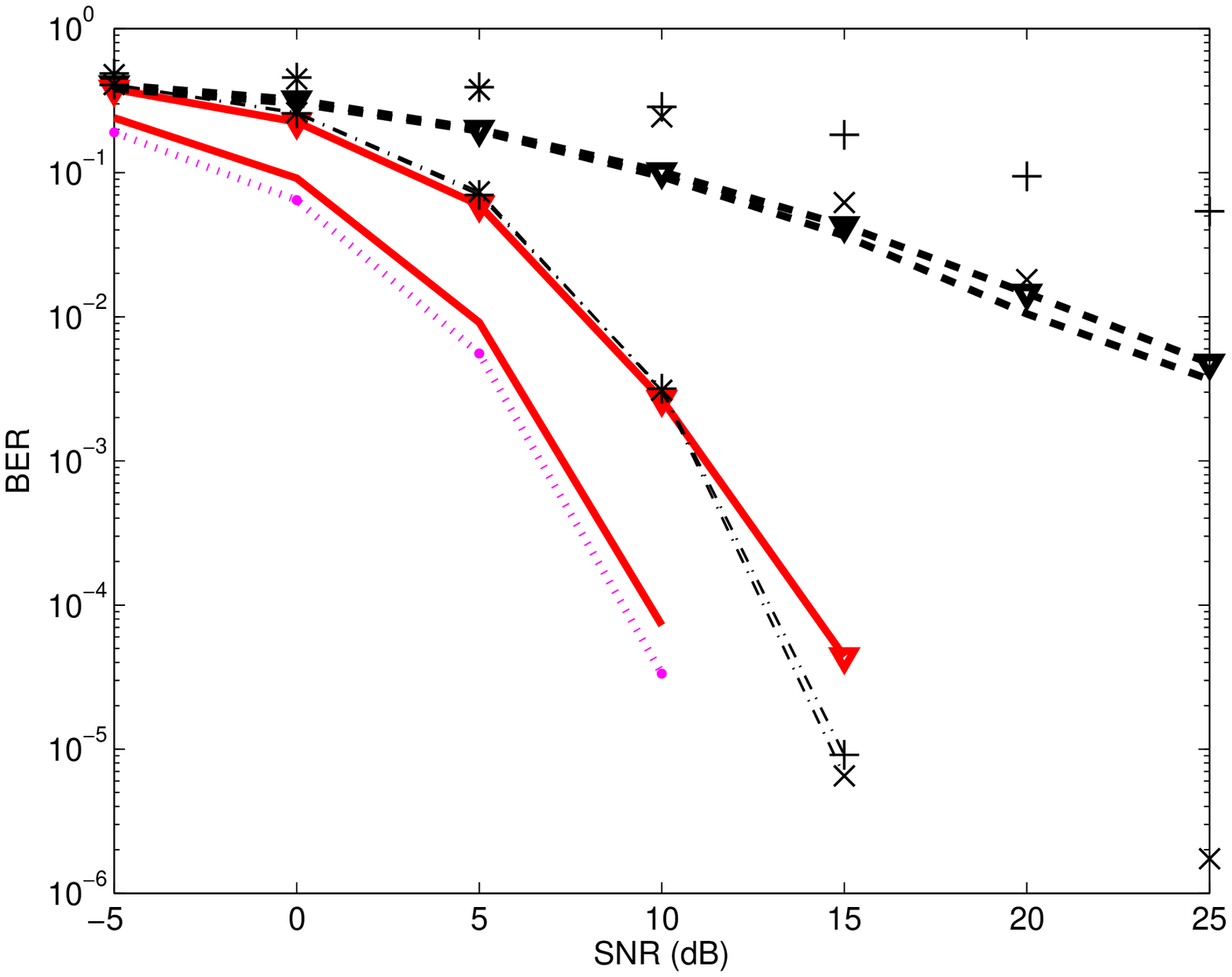}}}\vspace{1.0em}
\centerline{ \SetLabels
\L(0.34*-0.05) \footnotesize (b) 16-QAM modulation \\
\endSetLabels
\leavevmode \strut\AffixLabels{
\includegraphics[scale=0.40]{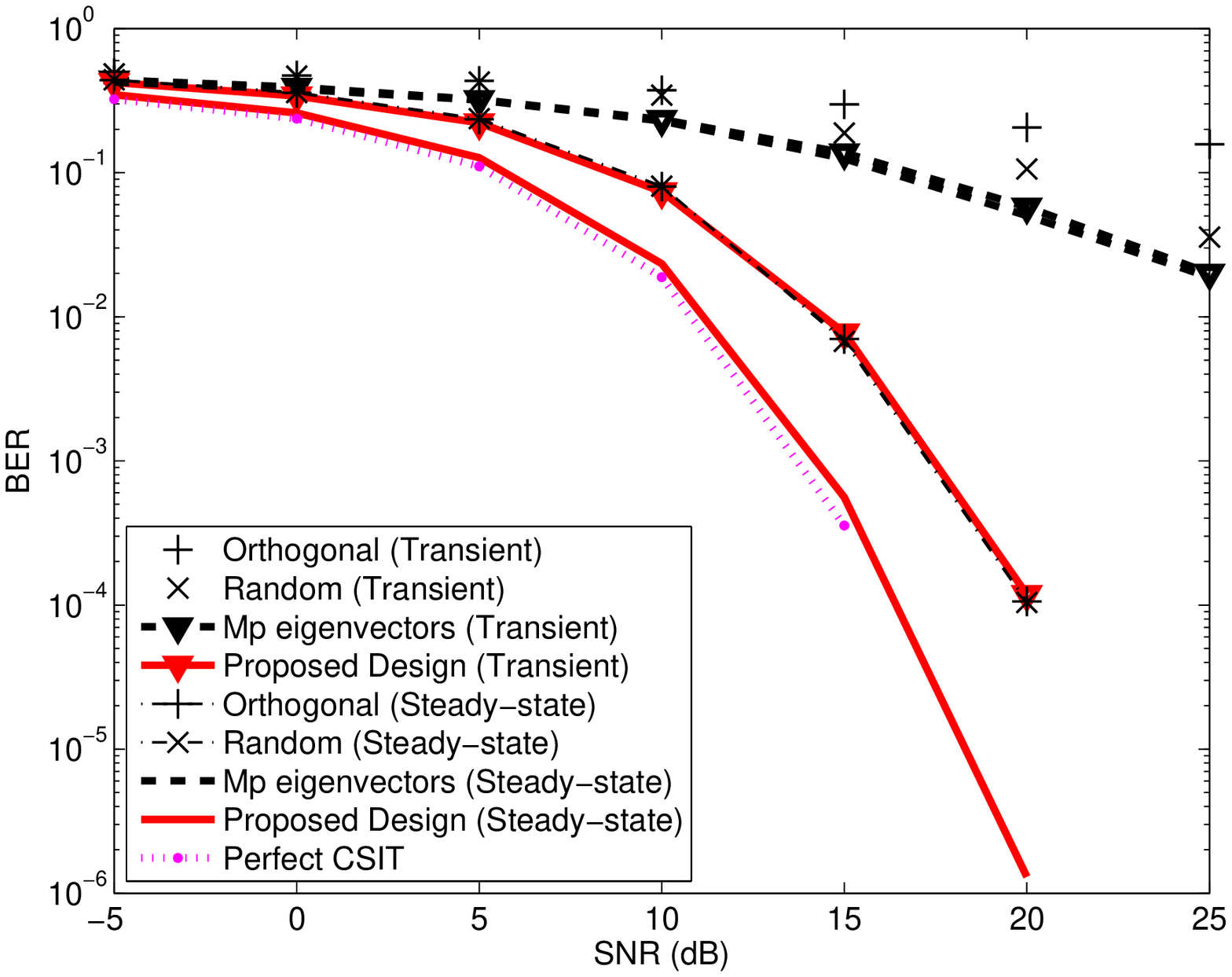}}}
\vspace{0.1em}
\caption{BER performance where $M=5$, $M_p=1$, and $v=3km/h$}
\label{fig:ber_planar}\vspace{-1.5em}
\end{figure}

Fig. \ref{fig:nmsesnr_planar_snr51015db_3kmh} shows the
performance of the two proposed algorithms for the considered one-ring
channel model: one with fixed pilot
power and the other with pilot power design. It is seen that proper power allocation can enhance the channel estimation performance especially both in low SNR and initial tracking periods, but the
performance gain is small and the two methods yield almost the
same performance at the steady state.
Thus, simpler Algorithm
\ref{alg:optimalMIMO} with fixed pilot power can be used without much performance loss.

Fig. \ref{fig:nmsesnr_planar_snr15db_3kmh} shows the channel
estimation performance of several pilot pattern design methods for
the considered one-ring model.  
{
It is seen that the proposed
method (Algorithm 1) significantly outperforms other pilot design
methods both in the transient and steady-state behaviors. 
{Especially, the proposed method yields a received SNR loss of approximately 3dB
compared to the perfect channel state information case during the transient tracking phase.
Orthogonal and random pilot beam patterns are
ineffective since they span all the $N_t$-dimensional space and
such patterns cannot capture the dominant channel uncertainty in
space at each pilot symbol time\cite{Noh&Zoltowski&Sung&Love:13ASILOMAR}.}
The fixed $M_p$ eigen-direction
method outperforms the random or orthogonal pilot design methods
in the beginning. This is because the estimated channel from the
fixed $M_p$ eigen-direction pilot design is a linear combination
of the fixed $M_p$ eigen-directions, and the use of this channel
estimate  as the beamforming direction yields a rough channel
matching in the begining. However, as time goes, the channel
estimation in the limited subspace is not enough for accurate
channel estimation, and this yields the performance saturation.
To assess the actual system performance loss due to channel estimation error, we investigated the bit error rate (BER) performance. Fig. \ref{fig:ber_planar} shows the BER performance based on the estimated channel corresponding to Fig.  \ref{fig:nmsesnr_planar_snr15db_3kmh} for the same setup. It is seen that
the proposed method significantly outperforms other methods. Note that the channel MSE performance directly affects on the BER performance.
}

\begin{figure*}[!t]
\centerline{ \SetLabels
\L(0.159*-0.02) \footnotesize (a) Channel estimation \\
\L(0.712*-0.02) \footnotesize (b) Received SNR \\
\endSetLabels
\leavevmode \strut\AffixLabels{
\includegraphics[scale=0.43]{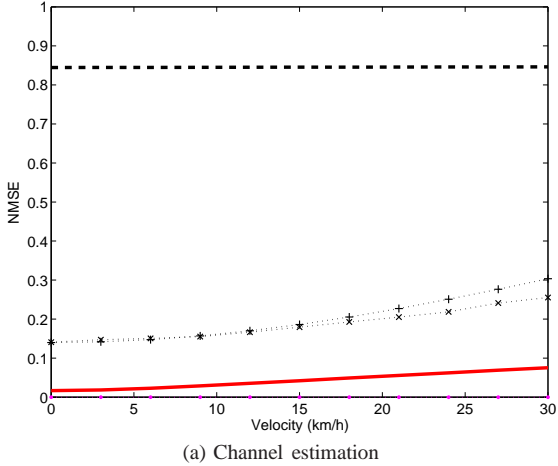}\hspace{1.5cm}
\includegraphics[scale=0.43]{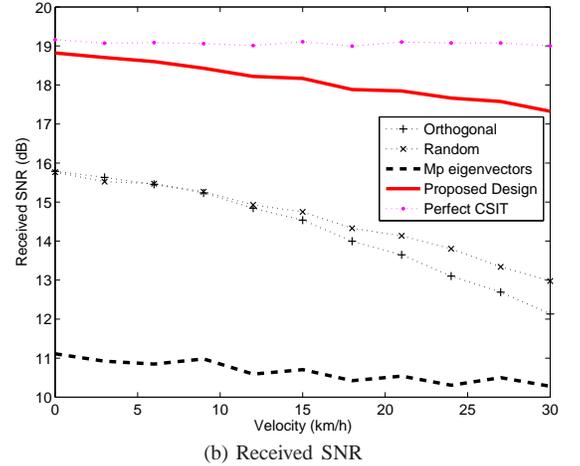}}}
\vspace{-0.1em}
\caption{NMSE and SNR versus the terminal velocity $v$ where $M=2$, $M_p=1$, and $\sigma_w^2=10^{-1.5}$}
\label{fig:nmsesnr_planar_snr20db_velocity}
\vspace{-1.7em}
\end{figure*}

We also investigated the performance variation due to the mobile speed. Fig. \ref{fig:nmsesnr_planar_snr20db_velocity} shows the
steady-state performance of several pilot beam pattern design
methods and the corresponding Kalman filtering channel estimation
channel as the mobile velocity $v$ varies from $0 km/h$ to $30
km/h$. Note that the proposed design yields much better
performance in the case of fast-fading when compared to the other
design methods.

\begin{figure}[!t]
\centerline{\includegraphics[scale=0.46]{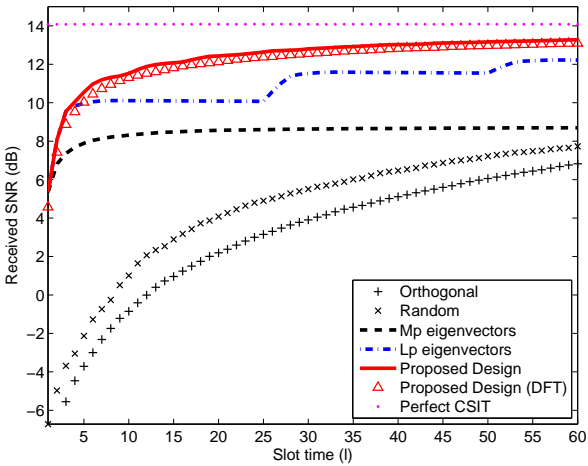}}
\vspace{-1.4em} \caption{Received SNR versus slot index $l$ where $M=5$, $M_p=2$, $\sigma_w^2=10^{-1}$, and $v=3km/h$}
\label{fig:nmsesnr_planar_snr10db_3kmh_blkfading} \vspace{-1.3em}
\end{figure}

{Finally, we evaluated the proposed design in the considered one-ring model using the $\Rbf_t$ estimation method based on the DFT matrix and the Toeplitz distribution theorem (TDT) presented in Section \ref{sec:discussion}.
Fig. \ref{fig:nmsesnr_planar_snr10db_3kmh_blkfading} shows the received SNR performance. (Here, we used the block-fading channel Gauss-Markov model in Section  \ref{subsec:blockading} since this case was not covered so far, but the performance is not much different from  the same for the symbol fading case.) We assumed that AoA and $\Delta$ are known. It is seen that the DFT/TDT-based method yields almost the same performance as the proposed algorithm with  perfectly known $\Rbf_\hbf$! Thus, the simple practical estimation of $\Rbf_\hbf$ based on the DFT and the TDT seems to work well. Here, to overcome the drawback of the method of using the fixed $M_p$ dominant eigenvectors of $\Rbf_\hbf$, we also considered a modified method that initially chooses $L_p ~(>M_p)$ dominant eigenvectors of $250 \times 250$ $\Rbf_\hbf$ and uses $M_p$ patterns out of the chosen $L_p$ patterns in a round-robin manner.  $L_p=50$ was used for Fig. \ref{fig:nmsesnr_planar_snr10db_3kmh_blkfading}. Note that up to the first 5 slots the modified method almost tracks the proposed method. This means that roughly 10 eigen-directions out of $L_p=$ 50 are most significant and contain most of the channel power. Hence, if $L_p$ were 10, the performance of the modified method should be very good and be comparable to that of the proposed method. However, the problem here is that one does not know the number of dominant eigen-directions containing most of the channel power {\it a priori} with a proper threshold level.  One can view that the proposed algorithm exploits both the most significant eigen-direction and the channel power of each direction over time.
}

\vspace{-0.3em}
\section{Conclusions} \label{sec:conclusion}

We have considered the problem of pilot beam pattern design for
massive MIMO systems, and  proposed a new method for pilot beam
pattern design for massive MIMO systems, based on the stationary
Gauss-Markov channel model, by exploiting  channel statistics such
as temporal and spatial channel correlation that can be used for
better system performance.  The proposed method yields a greedy
(i.e., sequentially optimal) sequence of pilot beam patterns with
low computational complexity by exploiting the properties of the
Kalman filtering and prediction error covariance matrices.
Furthermore, we have considered the joint design problem of pilot
beam pattern and pilot beam power and the extension of the
proposed method to the case of the block Gauss-Markov channel
model. Numerical results have validated the effectiveness of the
proposed algorithm, and it is shown that  the proposed pilot
design method significantly outperforms other pilot design methods
especially under the realistic one-ring channel correlation model.

\vspace{-0.3em}
\appendix

\subsection{Proof of Proposition \ref{pro:argminMMSE_mimo}}\label{app:argminMMSE_mimo}

See
\cite{Noh&Zoltowski&Sung&Love:13ASILOMAR} for the MISO case. We
here prove the MIMO case.

 {\it Case 1) $k \ne l M + 1$:} ~From
\eqref{eq:mesurementupdateP}, $\mathop{\arg
\min}_{\sbf_k}\mbox{tr}(\Pbf_{k|k})$ can be written as
\begin{equation}
\mathop{\arg \max}_{\sbf_k}~\text{tr}\left(\Pbf_{k|k-1}\Sbf_k(\Sbf_k^H\Pbf_{k|k-1}\Sbf_k+\sigma_w^2\Ibf_{N_r})^{-1}\Sbf_k^H\Pbf_{k|k-1}\right). \label{eq:objftntracePkkMIMO_v2}
\end{equation}
Since $\text{tr}(\Abf\Bbf\Cbf)=\text{tr}(\Bbf\Cbf\Abf)$ and
$\Sbf_k = \sbf_k \otimes \Ibf_{N_r}$, the cost function in
\eqref{eq:objftntracePkkMIMO_v2} can be rewritten as
\begin{align}
J&=\text{tr}\left(\left((\sbf_k\otimes\Ibf_{N_r})^H\Pbf_{k|k-1}(\sbf_k\otimes\Ibf_{N_r})+\sigma_w^2\Ibf_{N_r}\right)^{-1}\right. \nonumber \\
&~~~\left.(\sbf_k\otimes\Ibf_{N_r})^H\Pbf_{k|k-1}^2(\sbf_k\otimes\Ibf_{N_r})\right).\label{eq:objftntracePkkMIMO_v3}
\end{align}
Since the Kalman prediction error covariance matrix
$\Pbf_{k|k-1}=(\Ubf\otimes\Vbf)\Lambdabf(\Ubf\otimes\Vbf)^H$ by
the assumption, where
$\Lambdabf=\mbox{diag}(\Lambdabf_1,\cdots,\Lambdabf_{N_t})$,
$\Ubf\in\mathbb{C}^{N_t\times N_t }$ and
$\Vbf\in\mathbb{C}^{N_r\times N_r}$, and since the columns of
$\Ubf=[\ubf_1,\cdots,\ubf_{N_t}]$ span ${\mathbb C}^{N_t}$, we
have $\sbf_k = \sum_{i=1}^{N_t} c_i \ubf_i$, where $\sum_i |c_i|^2
= \rho_p$, and \eqref{eq:objftntracePkkMIMO_v3} can be rewritten
as
\begin{align}
J&=\text{tr}\left\{\bigl(\bigl[\Ubf^H\sbf_k\otimes\Vbf^H\bigr]^H\Lambdabf \bigl[\Ubf^H\sbf_k\otimes\Vbf^H\bigr]+\sigma_w^2\Ibf_{N_r}\bigr)^{-1} \right.\nonumber \\
&~~\left. \bigl(\Ubf^H\sbf_k\otimes\Vbf^H\bigr)^H\Lambdabf^2 \bigl(\Ubf^H\sbf_k\otimes\Vbf^H\bigr)\right\} \nonumber\\
&=
\text{tr}\left\{\left( \bigl[({\textstyle \sum_i} c_i \ebf_i)\otimes\Vbf^H\bigr]^H\Lambdabf \bigl[({\textstyle \sum_i} c_i \ebf_i)\otimes\Vbf^H\bigr] + \sigma_w^2\Ibf_{N_r}\right)^{-1} \right.\nonumber \\
&~~~~\left. \bigl(({\textstyle\sum_i} c_i \ebf_i)\otimes\Vbf^H\bigr)^H\Lambdabf^2\bigl(({\textstyle\sum_i} c_i \ebf_i)\otimes\Vbf^H\bigr)\right\} \label{eq:objftntracePkkMIMO_v4}\\
&= \text{tr}\left\{ \bigl({\textstyle\sum_i} |c_i|^2
\Lambdabf_i+\sigma_w^2\Ibf_{N_r}\bigr)^{-1} \bigl({\textstyle\sum_i} |c_i|^2
\Lambdabf_i^2\bigr)\right\},\label{eq:objftntracePkkMIMO_v44}
\end{align}
where $\ebf_i$ is the $i$-th unit vector, and the last step
\eqref{eq:objftntracePkkMIMO_v44} holds because
\begin{align}
\bigl(\ebf_i\otimes\Vbf^H\bigr)^H\Lambdabf^{p}\bigl(\ebf_j\otimes\Vbf^H\bigr) &=
\delta_{ij} \Vbf \Lambdabf_i^p\Vbf^H,
\label{eq:objftntracePkkMIMO_v5}
\end{align}
where $p\in\{1,2\}$ and $\delta_{ij}$ is the Kronecker delta.
 The
cost function \eqref{eq:objftntracePkkMIMO_v44} can be rewritten
as
\begin{equation} \label{eq:Prop1CostSumForm}
J(c_1,\cdots,c_{N_t}) = \sum_{j=1}^{N_r} \frac{\sum_m |c_m|^2
\lambda_{mj}^2}{\sum_n |c_n|^2 \lambda_{nj}+\sigma_w^2},
\end{equation}
where $\Lambdabf_i=
\mbox{diag}(\lambda_{i1},\cdots,\lambda_{iN_r})$. The Lagrangian
of the optimization of \eqref{eq:Prop1CostSumForm} is given by
\[
\Lc = \sum_{j=1}^{N_r} \frac{\sum_m |c_m|^2 \lambda_{mj}^2}{\sum_n
|c_n|^2 \lambda_{nj}+\sigma_w^2}  + \nu\left(\sum_m |c_m|^2 -
\rho_p\right),
\]
where $\nu$ is a  Lagrange dual variable. The Karush-Kuhn-Tucker
(KKT) conditions of the optimization of
\eqref{eq:Prop1CostSumForm} are given by {\footnotesize
\begin{align*}
0&= \frac{\partial \Lc}{\partial c_i^*}
= \sum_j \frac{c_i\lambda_{ij}^2(\sum_n |c_n|^2
\lambda_{nj}+\sigma_w^2)-c_i\lambda_{ij}(\sum_m |c_m|^2
\lambda_{mj}^2)}{(\sum_n |c_n|^2 \lambda_{nj}+\sigma_w^2)^2} +\nu
c_i.
\end{align*}}\noindent
It is easy to verify that $c_{i^\prime} =\rho_p e^{\iota\theta}$
for some $i^\prime\in \{1,2,\cdots,N_t\}$ and $c_i =0$ for all
$i\ne i^\prime$ with $\nu=-\sum_j \frac{\lambda_{i^\prime
j}^2\sigma_w^2}{(\rho_p\lambda_{i^\prime j}^2)^2}$ satisfies the
KKT conditions. Since \eqref{eq:Prop1CostSumForm} is not convex in
terms of $\{c_i\}$, the solution to the KKT conditions is not
unique. However, all such solutions with only one non-zero $c_i$
are stationary points of the optimization, i.e., each of them is a
local optimum. Among such solutions the best one is given by $c_i
= \sqrt{\rho_p}$ for $i=i_k$ and $c_i =0$ for all $i\ne i_k$,
where
\begin{align}
i_k:=\argmax_{i}&~
\text{tr}\left\{\bigl(\rho_p\Lambdabf_i+\sigma_w^2\Ibf_{N_r}\bigr)^{-1}
\bigl(\rho_p \Lambdabf_i^2\bigr)\right\} \nonumber \\
=\argmax_{i}&~
\sum_{j=1}^{N_r}\frac{\rho_p\lambda_{ij}^2}{\rho_p\lambda_{ij}+\sigma_w^2},\label{eq:objftntracePkkMIMO_v7}
\end{align}
and $\sbf_k = \sqrt{\rho_p} \ubf_{i_k}$ is a locally optimal
solution to minimizing $\mbox{tr}(\Pbf_{k|k})$.

{\it Case 2) $k = l M + 1$:} In this case, we have $M_d$
prediction steps without a measurement update step before the
first pilot symbol time $k$ in the $l$-th slot. In this case,
still the measurement update form \eqref{eq:mesurementupdateP} at
$k$ is valid with $\Pbf_{k|k-1}$ replaced by the error covariance
matrix  $\Pbf_{k|(l -1)M+M_p}$ of  the Kalman prediction for time
$k$ based on all the previous pilot beam patterns.  Hence, the
proof in Case 1) is applicable to this case just with
$\Pbf_{k|k-1}$ replaced by $\Pbf_{k|(l -1)M+M_p}$.
$\hfill{\blacksquare}$ \vspace{0.5em}

\subsection{Derivation of $\Pbf_{k|k}$}\label{subsec:equations_for_P_kk}
\vspace{-1.5em}
{\small
\begin{align}
&\Pbf_{k|k} \nonumber\\
&=\Pbf_{k|k-1} - \Pbf_{k|k-1}\Sbf_k (\Sbf_k^H\Pbf_{k|k-1}\Sbf_k+\sigma_w^2\mathbf{I}_{N_r})^{-1} \Sbf_k^H\Pbf_{k|k-1} \nonumber\\
&=(\Ubf\otimes\Vbf)\Lambdabf^{(k)}(\Ubf\otimes\Vbf)^H -
    (\Ubf\otimes\Vbf)\Lambdabf^{(k)}(\sqrt{\rho_p}\ebf_{i_k}\otimes\Vbf^H) \nonumber\\
&~~~\left[  (\sqrt{\rho_p}\ebf_{i_k}\otimes\Vbf^H)^H\Lambdabf^{(k)}(\sqrt{\rho_p}\ebf_{i_k}\otimes\Vbf^H)+\sigma_w^2\Ibf_{N_r} \right]^{-1} \nonumber \\
&~~~~(\sqrt{\rho_p}\ebf_{i_k}\otimes\Vbf^H)^H\Lambdabf^{(k)}(\Ubf\otimes\Vbf)^H
\\
 &\stackrel{(a)}{=}
(\Ubf\otimes\Vbf)\Lambdabf^{(k)}(\Ubf\otimes\Vbf)^H -
(\Ubf\otimes\Vbf)\Lambdabf^{(k)}(\sqrt{\rho_p}\ebf_{i_k}\otimes\Vbf^H) \nonumber\\
  &~~~~\left[\Vbf\bigl(\rho_p \Lambdabf^{(k)}_{i_k}+\sigma_w^2\Ibf_{N_r}\bigr)^{-1}\Vbf^H\right] (\sqrt{\rho_p}\ebf_{i_k}\otimes\Vbf^H)^H \nonumber \\
  &~~~~~\Lambdabf^{(k)}(\Ubf\otimes\Vbf)^H\nonumber \\
&\stackrel{(b)}{=}
(\Ubf\otimes\Vbf)\Lambdabf^{(k)}(\Ubf\otimes\Vbf)^H -   (\Ubf\otimes\Vbf)\Lambdabf^{(k)} \nonumber \\
&~~~~\left[\rho_p(\ebf_{i_k}\ebf_{i_k}^T)\otimes \bigl(\rho_p \Lambdabf^{(k)}_{i_k}+\sigma_w^2\Ibf_{N_r}\bigr)^{-1}\right]\Lambdabf^{(k)}(\Ubf\otimes\Vbf)^H\nonumber \\
&\stackrel{(c)}{=}
(\Ubf\otimes\Vbf)\Lambdabf^{(k)}(\Ubf\otimes\Vbf)^H -   (\Ubf\otimes\Vbf)    \left\{(\ebf_{i_k}\ebf_{i_k}^T)\otimes \right. \nonumber \\
&~~~~\left. \left[\rho_p\Lambdabf^{(k)}_{i_k}\bigl(\rho_p\Lambdabf^{(k)}_{i_k}+\sigma_w^2\Ibf_{N_r}\bigr)^{-1}\Lambdabf^{(k)}_{i_k}\right]\right\}(\Ubf\otimes\Vbf)^H\nonumber \\
&=
(\Ubf\otimes\Vbf)\left\{ \Lambdabf^{(k)} - (\ebf_{i_k}\ebf_{i_k}^T)\otimes \right.\nonumber\\
&~~~\left. \left[\rho_p\Lambdabf^{(k)}_{i_k}\bigl(\rho_p\Lambdabf^{(k)}_{i_k}+\sigma_w^2\Ibf_{N_r}\bigr)^{-1}\Lambdabf^{(k)}_{i_k}\right]\right\}(\Ubf\otimes\Vbf)^H, \nonumber
\end{align}}
where the equality $(a)$ follows because
\begin{align}
&\left[\bigl(\sqrt{\rho_p}\ebf_{i_k}\otimes\Vbf^H\bigr)^H\Lambdabf^{(k)} \bigl(\sqrt{\rho_p}\ebf_{i_k}\otimes\Vbf^H\bigr)+\sigma_w^2\Ibf_{N_r}\right]^{-1} \nonumber\\
&=\Vbf\left(\rho_p \Lambdabf^{(k)}_{i_k}+\sigma_w^2\Ibf_{N_r}\right)^{-1}\Vbf^H,
\end{align}
and the equality $(b)$ follows because
$(\Abf_1\Abf_2)\otimes(\Bbf_1\Bbf_2)=(\Abf_1\otimes\Bbf_1)(\Abf_2\otimes\Bbf_2)$.
The equality $(c)$ holds because
\begin{align}
&\Lambdabf^{(k)}\left[\rho_p(\ebf_{i_k}\ebf_{i_k}^T)\otimes \bigl(\rho_p \Lambdabf^{(k)}_{i_k}+\sigma_w^2\Ibf_{N_r}\bigr)^{-1}\right]\Lambdabf^{(k)} \nonumber \\
&=
(\ebf_{i_k}\ebf_{i_k}^T)\otimes  \left[\rho_p\Lambdabf^{(k)}_{i_k}\bigl(\rho_p\Lambdabf^{(k)}_{i_k}+\sigma_w^2\Ibf_{N_r}\bigr)^{-1}\Lambdabf^{(k)}_{i_k}\right]. \nonumber
\end{align}

\subsection{Proof of Proposition \ref{prop:powerallocation}}\label{app:powerallocation}
For the $l$-th pilot transmission period with $k=l
M+m$, let $\rhobf=[\rho_{l M+1},\ldots,\rho_{l M+M_p}]^T$ be a
power allocation vector with the pilot beam pattern sequence
determined by $\{\Kc_i$, $1\le i \le N_t\}$.  The channel
estimation MSE at time $l M+M_p$ is given by
\begin{align}
&\text{tr}(\Pbf_{l M+M_p| l M+M_p}) =
\sum_{i=1}^{N_t} \text{tr}(\bar{\Lambdabf}^{(l M+M_p)}_i) \nonumber \\
&= \sum_{i=1}^{N_t}
\text{tr}\left(a^{2(\bar{p}_i-1)}\bar{\Lambdabf}^{(k_{|\Kc_i|}^i)}_i
+ (1-a^{2(\bar{p}_i-1)}) \Lambdabf^{(1)}_i\right),
\label{eq:mseEndPilot}
\end{align}
where $\bar{\Lambdabf}^{(k)}_i\in\mathbb{R}^{N_r\times N_r}$ is
the $i$-th diagonal sub-block of $\bar{\Lambdabf}^{(k)}$ defined
\eqref{eq:measurementUpdateReductionMIMO}. ($\bar{p}_i = M_p+1$,
$k_{|\Kc_i|}^i=lM$ when $|\Kc_i|=0$.) \eqref{eq:mseEndPilot} holds
because $\ubf_i$ only affects the $i$-th subblock of the
eigenvalue matrix and the MSE  for the $i$-th block  at the end of
the pilot period is given by channel prediction from the last
pilot use of $\ubf_i$ at time $k_{|\Kc_i|}^i$. Combining Kalman
prediction and measurement update steps, we have  for each
$k^i_j\in \Kc_i$
 {\small \begin{align}
&\text{tr}(\bar{\Lambdabf}^{(k^i_j)}_i) \nonumber \\
&=
\text{tr}\left(\frac{\sigma_w^2\bigl(a^{2p_{j}^i}\bar{\Lambdabf}^{(k^i_{j-1})}_i
+ (1-a^{2p_{j}^i})\Lambdabf^{(1)}_i\bigr)}{\rho_{
k^i_j}\bigl(a^{2p_{j}^i}\bar{\Lambdabf}^{(k^i_{j-1})}_i +
(1-a^{2p_{j}^i})\Lambdabf^{(1)}_i\bigr)+\sigma_w^2\Ibf_{N_r}}\right)
\label{eq:optimalpowerallocation} \\
&=f\bigl(\bar{\Lambdabf}^{(k^i_{j-1})}_{i,0}\bigr)
\end{align}}\noindent
where $f\bigl(\bar{\Lambdabf}^{(k^i_{j-1})}_{i,\epsilon}\bigr)$
and  $\bar{\Lambdabf}^{(k^i_{j-1})}_{i,\epsilon}$  are defined in
(\ref{eq:vardef_1}, \ref{eq:vardef_3}).  (Here, we have slight
abuse of notation. $\Abf/\Bbf$ means $\Bbf^{-1}\Abf$ for two
matrices $\Abf$ and $\Bbf$.) Proof is by an iterative argument. We
start from $j=|\Kc_i|$ and $j-1=|\Kc_i|-1$ for the original
$\Kc_i$. By Lemma \ref{lem:powerallocation} and Remark
\ref{rem:powerallocation}, \eqref{eq:optimalpowerallocation} is
reduced by updating $\tilde{\rho}_{k_j^i}=\rho_{k_j^i}+\rho_{
k_{j-1}^i}$ and $\tilde{\rho}_{k_{j-1}^i}=0$, when we consider the
two power values for $j-1$ and $j$. With this improvement, we
construct a new
$\Kc_i^\prime=\{k_1^i,\cdots,k_{j-2}^i,k^i_{|\Kc_i|}\}$ with
$|\Kc_i^\prime|=|\Kc_i|-1$ and a new power allocation  $[\rho_{
k_1^i},\cdots,\rho_{ k_{|\Kc_i|-2}^i},\rho_{
k_{|\Kc_i|-1}^i}+\rho_{ k_{|\Kc_i|}^i}]^T$ for $\Kc_i^\prime$.
Then, we apply the same argument to the last two power terms of
the newly constructed $\Kc_i^\prime$. In this way,
\eqref{eq:optimalpowerallocation} is minimized by allocating all
the power for the $i$-th eigen-direction to $k^i_{|\Kc_i|}$ for
the original $\Kc_i$. Since \eqref{eq:mseEndPilot} is a monotone
increasing function of $\text{tr}(\bar{\Lambdabf}^{(k^i_j)}_i)$,
we have the claim. $\hfill{\blacksquare}$

\begin{lemma}\label{lem:powerallocation}
Given any $\rho_{k_{j-1}^i}, \rho_{k_j^i} \in\mathbb{R}_+$, set
$\tilde{\rho}_{k_{j-1}^i}=\rho_{k_{j-1}^i}-\epsilon$ and
$\tilde{\rho}_{k_j^i}=\rho_{k_j^i}+\epsilon$  for any $\epsilon
\in [0, \rho_{k_{j-1}^i}]$. Then,  the following holds:
\begin{align}
f\bigl(\bar{\Lambdabf}^{(k_{j-1}^i)}_{i,0}\bigr) -
f\bigl(\bar{\Lambdabf}^{(k_{j-1}^i)}_{i,\epsilon}\bigr) \ge 0,
\label{eq:ineqforpower}
\end{align}
where {\small
\begin{align}
f\bigl(\bar{\Lambdabf}^{(k_{j-1}^i)}_{i,\epsilon}\bigr) &=
\text{tr}\left(
\frac{\sigma_w^2\bigl(a^{2p_{j}^i}\bar{\Lambdabf}^{(k_{j-1}^i)}_{i,\epsilon}
+ (1-a^{2p_{j}^i})\Lambdabf^{(1)}_i\bigr)}
{\tilde{\rho}_{k_j^i}\bigl(a^{2p_{j}^i}\bar{\Lambdabf}^{(k_{j-1}^i)}_{i,\epsilon}
+
(1-a^{2p_{j}^i})\Lambdabf^{(1)}_i\bigr)+\sigma_w^2\Ibf_{N_r}}\right)
\label{eq:vardef_1}\\
 \bar{\Lambdabf}^{(k_{j-1}^i)}_{i,\epsilon} &=
\frac{\sigma_w^2 \Lambdabf^{(k_{j-1}^i)}_i
}{\tilde{\rho}_{k_{j-1}^i}\Lambdabf^{(k_{j-1}^i)}_i
+\sigma_w^2\Ibf_{N_r}} \label{eq:vardef_3}\\
\Lambdabf^{(k_{j-1}^i)}_i &=
a^{2p_{j-1}^i}\bar{\Lambdabf}^{(k_{j-2}^i)}_i + (1-a^{2p_{j-1}^i})\Lambdabf^{(1)}_i  \\
k_0^i&=l M,~k_j^i\in\Kc_i,~\text{and } 2\le j\le |\Kc_i|.
\label{eq:vardef_4}
\end{align}}
\end{lemma}
\vspace{-0.2em}

{\em Proof:} For notational simplicity, we omit the upper index
$i$ of $k_j^i$ and $p_j^i$ when there is no ambiguity.  Define
$\tilde{\Dbf}:=a^{2p_{j}}\bar{\Lambdabf}^{(k_{j-1})}_{i,\epsilon}+(1-a^{2p_{j}})\Lambdabf^{(1)}_i$
and
$\Dbf:=a^{2p_{j}}\bar{\Lambdabf}^{(k_{j-1})}_i+(1-a^{2p_{j}})\Lambdabf^{(1)}_i$
with
$\bar{\Lambdabf}^{(k_{j-1})}_i:=\bar{\Lambdabf}^{(k_{j-1})}_{i,\epsilon}|_{\epsilon=0}$.
Then, \eqref{eq:ineqforpower} can be rewritten as
{\small
\begin{align}
&\text{tr}\left(\frac{\sigma_w^2\Dbf}{\rho_{k_j} \Dbf + \sigma_w^2\Ibf_{N_r}} - \frac{\sigma_w^2\tilde{\Dbf}}{\tilde{\rho}_{k_j} \tilde{\Dbf} + \sigma_w^2\Ibf_{N_r}}\right) \nonumber\\
&= \text{tr}\left(\frac{\sigma_w^2\bigl(\epsilon \Dbf\tilde{\Dbf}
+ \sigma_w^2( \Dbf- \tilde{\Dbf})\bigr)}{(\rho_{k_j} \Dbf +
\sigma_w^2\Ibf_{N_r})(\tilde{\rho}_{k_j} \tilde{\Dbf} +
\sigma_w^2\Ibf_{N_r})}\right).\label{eq:eq:ineqforpower_v2}
\end{align}}
Note that the denominator of the right-hand side (RHS) in
\eqref{eq:eq:ineqforpower_v2} is obviously positive definite and
the numerator is also positive semi-definite because each term on
the RHS in \eqref{eq:ineqforpower_v3} is positive semi-definite
because
\begin{align}
&\epsilon \Dbf\tilde{\Dbf} + \sigma_w^2( \Dbf- \tilde{\Dbf}) \nonumber\\
&= \epsilon(1-a^{2p_{j}})^2(\Lambdabf^{(1)}_i)^2 +
 \epsilon\frac{a^{2p_{j}}(1-a^{2p_{j}})\sigma_w^2\Lambdabf^{(k_{j-1})}_i}{(\rho_{k_j} \Dbf + \sigma_w^2\Ibf_{N_r})(\tilde{\rho}_{k_j} \tilde{\Dbf} + \sigma_w^2\Ibf_{N_r})}\nonumber\\
&~~~
\left[(2\rho_{k_{j-1}}-\epsilon)\Lambdabf^{(1)}_i\Lambdabf^{(k_{j-1})}_i+ \right.\nonumber\\
&~~~\left.
\sigma_w^2\bigl(\Lambdabf^{(1)}_i+a^{2p_{j-1}}(\Lambdabf^{(1)}_i-\bar{\Lambdabf}^{(k_{j-2})}_i)\bigr)\right].
\label{eq:ineqforpower_v3}
\end{align}
Note that $\Lambdabf_i^{(1)}  \succeq \bar{\Lambdabf}_i^{(k)}$ for
all $k$. (Remember that the channel is stationary and the
measurement update only improves the channel estimation quality.)
Hence, we have the claim. $\hfill{\blacksquare}$

\begin{remark}\label{rem:powerallocation} In case that we control $\rho_{k_{j-1}}, \rho_{k_j}
\in\mathbb{R}_+$,
$f\bigl(\bar{\Lambdabf}^{(k_{j-1})}_{i,\epsilon}\bigr)$ is
minimized when $\epsilon=\epsilon^\prime:=\rho_{k_{j-1}}$.
 This can easily be shown
by $f\bigl(\bar{\Lambdabf}^{(k_{j-1})}_{i,\epsilon}\bigr) -
f\bigl(\bar{\Lambdabf}^{(k_{j-1})}_{i,\epsilon^\prime}\bigr) \ge
0$. One can write a similar equation to
\eqref{eq:eq:ineqforpower_v2}. Although the detail is not shown
here, in this case the corresponding denominator is positive
definite and the corresponding numerator includes obviously
positive semi-definite term and the term
\begin{align}
&(\rho_{k_{j-1}}- \epsilon)(1-a^{2p_{j}})
\left[\Lambdabf^{(1)}_i\left((1-a^{2(p_{j-1}+p_{j})})\Lambdabf^{(1)}_i+  \right.\right.\nonumber\\
&\left.\left. a^{2(p_{j-1}+p_{j})}\bar{\Lambdabf}^{(k_j)}_i\right)
+
a^{2(p_{j-1}+p_{j})}(\Lambdabf^{(1)}_i-\bar{\Lambdabf}^{(k_j)}_i)\bar{\Lambdabf}^{(k_{j-1})}_{i,\epsilon}\right],
\nonumber 
\end{align}
which is positive semi-definite.
\end{remark}

\subsection{Proof of Proposition \ref{pro:argminMMSE_blkfading}}\label{app:argminMMSE_blkfading}
From \eqref{eq:mesurementupdateP} and \eqref{eq:objftntracePkkMIMO_v3},
$\argmin_{\Sbf_{l}}\text{tr}(\Pbf_{l | l})$ can be written as
\noindent
\begin{align}
\argmax_{\Sbf_{l}} ~\text{tr}\left([\Sbf_{l}^H\Pbf_{l
|l-1}\Sbf_{l}+\sigma_w^2\Ibf_{M_p}]^{-1}\Sbf_{l}^H\Pbf_{l
|l-1}^2\Sbf_{l}\right). \label{eq:objftn_blkSk}
\end{align}
For  orthogonal pilot signals, the objective function
\eqref{eq:objftn_blkSk} can be rewritten as
\begin{align}
\text{tr}\left([\Sbf_{l}^H(\Pbf_{l
|l-1}+{\sigma_w^2/\rho_p}\Ibf_{N_t})\Sbf_{l}]^{-1}\Sbf_{l}^H\Pbf_{l|l-1}^2\Sbf_{l}\right). \label{eq:objftn_blkSk_v2}
\end{align}
Define $\Pbf_{l,\sigma_w}:=\Pbf_{l
|l-1}+{\sigma_w^2/\rho_p}\Ibf_{N_t}=\Pbf_{l,\sigma_w}^{1/2}\Pbf_{l,\sigma_w}^{H/2}$
and $\Fbf:=\Pbf_{l,\sigma_w}^{H/2}\Sbf_{l}$. Then,
\eqref{eq:objftn_blkSk_v2} can be rewritten as
\begin{align}
&\text{tr}\left((\Fbf^H\Fbf)^{-1}\Fbf^H\Pbf_{l,\sigma_w}^{-1/2}\Pbf_{l |l-1}^2\Pbf_{l,\sigma_w}^{-H/2}\Fbf\right) \nonumber \\
&=
\text{tr}\left((\Fbf^H\Fbf)^{-H/2}\Fbf^H\Pbf_{l,\sigma_w}^{-1/2}\Pbf_{l |l-1}^2\Pbf_{l,\sigma_w}^{-H/2}\Fbf(\Fbf^H\Fbf)^{-1/2}\right) \label{eq:objftn_blkSk_v3}\\
&= \text{tr}\left(\Bbf^H\Pbf_{l,\sigma_w}^{-1/2}\Pbf_{l
|l-1}^2\Pbf_{l,\sigma_w}^{-H/2}\Bbf\right),
\label{eq:objftn_blkSk_v4}
\end{align}
where $\Bbf:=\Fbf(\Fbf^H\Fbf)^{-1/2}$. The equality
\eqref{eq:objftn_blkSk_v3} holds by the positive definiteness of
$\Fbf^H\Fbf$ and
$\text{tr}(\Abf\Bbf\Cbf)=\text{tr}(\Bbf\Cbf\Abf)$. Because
$\Bbf^H\Bbf={\Ibf_{M_p}}$, the optimal $\Bbf$ that maximizes
\eqref{eq:objftn_blkSk_v4} is given by the $M_p$ dominant
eigenvectors of $\Pbf_{l,\sigma_w}^{-1/2}\Pbf_{l
|l-1}^2\Pbf_{l,\sigma_w}^{-H/2}$ by Ky-Fan\cite{Fan:50NAS}. Let
the ED of $\Pbf_{l |l-1}$ be $\Pbf_{l
|l-1}=\Ubf\Lambdabf^{(l)}\Ubf^H$, where the diagonal matrix
$\Lambdabf^{(l)}$ contains the eigenvalues of $\Pbf_{l |l-1}$ in a
decreasing order. Then, $\Pbf_{l,\sigma_w}^{-1/2}\Pbf_{l
|l-1}^2\Pbf_{l,\sigma_w}^{-H/2}$ is given by
\begin{align}
\Pbf_{l,\sigma_w}^{-1/2}\Pbf_{l |l-1}^2\Pbf_{l,\sigma_w}^{-H/2} &=
\Ubf\left(\frac{(\Lambdabf^{(l)})^2}{\Lambdabf^{(l)} +
{\sigma_w^2/\rho_p}\Ibf_{N_t}}\right)\Ubf^H,
\end{align}
from $\Pbf_{l,\sigma_w} =
\Ubf(\Lambdabf^{(l)}+{\sigma_w^2/\rho_p}\Ibf_{N_t})\Ubf^H$. Since
$g(x)=\frac{x^2}{x+\sigma^2},~x\ge 0$ is a monotone increasing
function of $x$, $\Bbf={\Ubf(:,1:M_p)}$, which is
achieved by  $\Sbf_{l}=\sqrt{\rho_p}\Ubf(:,1:M_p)$.
\hfill{$\blacksquare$} \vspace{0.4em}

\subsection{Power Allocation}\label{subsec:poweralloation}

The problem of \eqref{eq:optprobPowerAllocation} can be solved by
the standard convex optimization method.  The Lagrangian of the
problem is given by
\begin{align}
L(\rhobf,\xibf,\nu)&=
\sum_{i:|\Kc_i|=1}\sum_{j=1}^{N_r} \frac{ a^{2(lM+M_p-k^i)}\sigma_w^2 \lambda^{(k^i)}_{ij}  }{\rho_{k^i}\lambda^{(k^i)}_{ij} + \sigma_w^2}  - \nonumber \\
&~~~\sum_{i:|\Kc_i|=1} \xi_{k^i}\rho_{k^i} +
\nu\left(\sum_{i:|\Kc_i|=1}\rho_{k^i} - M_p\rho_p\right),
\nonumber
\end{align}
where $\xi_{k^i}$ and $\nu$ are the Lagrange multipliers
associated to the constraints, and $\lambdabf^{(k^i)}_i=\text{diag}(\Lambdabf^{(k^i)}_i)$ for $k^i\in\Kc_i$. The Karush-Kuhn-Tucker (KKT) conditions are then
written as
\begin{align}
\rho_{k^i} &\ge 0,~~ \sum_{i:|\Kc_i|=1}\rho_{k^i}  = M_p \rho_p,  \label{eq:kktcond1}\\
\xi_{k^i}   &\ge 0,~ 
\xi_{k^i}\rho_{k^i} = 0, \label{eq:kktcond3} \\
\frac{\partial L(\boldsymbol{\rho},\boldsymbol{\xi},\nu)}{\partial
\rho_{k^i}} &=
-\sum_{j=1}^{N_r}\frac{a^{2(l M+M_p-k^i)}\sigma_w^2\bigl(\lambda^{(k^i)}_{ij}\bigr)^2}{\left(\rho_{k^i}\lambda^{(k^i)}_{ij}+\sigma_w^2\right)^2}
- \xi_{k^i} + \nu = 0.\nonumber
\end{align}
From the above conditions, we have
\begin{align}
\sum_{j=1}^{N_r}\frac{a^{2(l M+M_p-k^i)}\sigma_w^2\bigl(\lambda^{(k^i)}_{ij}\bigr)^2}{\left(\rho_{k^i}\lambda^{(k^i)}_{ij}
+\sigma_w^2\right)^2} &\le \nu, \label{eq:kktcond_v2}\\
\left(\nu -
\sum_{j=1}^{N_r}\frac{a^{2(lM+M_p-k^i)}\sigma_w^2\bigl(\lambda^{(k^i)}_{ij}\bigr)^2}{\left(\rho_{k^i}\lambda^{(k^i)}_{ij}+\sigma_w^2\right)^2}\right)
\rho_{k^i} &= 0. \label{eq:kktcond_v3}
\end{align}
 If $\frac{a^{2(lM+M_p-k^i)}}{\sigma_w^2}\sum_{j=1}^{N_r}\bigl(\lambda^{(k^i)}_{ij}\bigr)^2>\nu$,
\eqref{eq:kktcond_v2} holds only if $\rho_{k^i}>0$, and by
\eqref{eq:kktcond_v3} this implies  that 
{\small
\begin{equation}
\nu =
\sum_{j=1}^{N_r}\frac{a^{2(lM+M_p-k^i)}\sigma_w^2\bigl(\lambda^{(k^i)}_{ij}\bigr)^2}{\left(\rho_{k^i}\lambda^{(k^i)}_{ij}+\sigma_w^2\right)^2}.
\label{eq:OptSolutionNu}
\end{equation}}
 If $\frac{a^{2(lM+M_p-k^i)}}{\sigma_w^2}\sum_{j=1}^{N_r}\bigl(\lambda^{(k^i)}_{ij}\bigr)^2\le
\nu$, then $\rho_{k^i}=0$ because we have
{\footnotesize\begin{equation} \nu\ge
\frac{a^{2(lM+M_p-k^i)}}{\sigma_w^2}\sum_{j=1}^{N_r}\bigl(\lambda^{(k^i)}_{ij}\bigr)^2
>
\sum_{j=1}^{N_r}\frac{a^{2(lM+M_p-k^i)}\sigma_w^2\bigl(\lambda^{(k^i)}_{ij}\bigr)^2}{\left(\rho_{k^i}\lambda^{(k^i)}_{ij}+\sigma_w^2\right)^2}.
\nonumber
\end{equation}}\noindent
\eqref{eq:kktcond_v3} holds only if $\rho_{k^i}=0$.

When $N_r=1$, the optimal power allocation is determined from \eqref{eq:OptSolutionNu} as
\begin{align}
\rho_{k^i} = \left(a^{l M+M_p-k^i}\frac{\sigma_w}{\sqrt{\nu}} -
\frac{\sigma_w^2}{\lambda^{(k^i)}_i}\right)^+,
\end{align}
where $\Kc_i=\{k^i\}$ and $\nu$ is determined by the power
constraint \eqref{eq:kktcond1}, given by
\begin{equation}
\sqrt{\nu} = \sigma_w\frac{1-a^{M_p}}{1-a}\left(M_p\rho_p +
\sigma_w^2\sum_{i:|\Kc_i|=1}\frac{1}{\lambda^{(k^i)}_i}\right)^{-1}.
\end{equation}

\subsection{Suboptimal Power Allocation}\label{subsec:poweralloationsuboptimal}

Consider the high SNR case first, i.e.,
$\rho_{k^i}\lambda^{(k^i)}_{ij}\gg \sigma_w^2$, where
$\lambdabf^{(k^i)}_i=\text{diag}(\Lambdabf^{(k^i)}_i)$. The cost
function \eqref{eq:optprobPowerAllocation} can be written as
{\small\begin{align*}
&\sum_{i:|\Kc_i|=1} \text{tr}\left(\frac{ a^{2(l M+M_p-k^i)} \sigma_w^2 \Lambdabf^{(k^i)}_i }{\rho_{k^i}\Lambdabf^{(k^i)}_i + \sigma_w^2\Ibf_{N_r}}  \right) \\
&\simeq \sum_{i:|\Kc_i|=1} \text{tr}\left(\frac{ a^{2(lM+M_p-k^i)}
\sigma_w^2\Ibf_{N_r}   }{\rho_{k^i}\Ibf_{N_r}} \right) =
N_r\sigma_w^2 \sum_{i:|\Kc_i|=1} \frac{ a^{2(l M+M_p-k^i)} }{\rho_{k^i}},\\
&\Rightarrow~ \min_{\boldsymbol\rho} \sum_{i:|\Kc_i|=1} \frac{
a^{2(l M+M_p-k^i)} }{\rho_{k^i}}.
\label{eq:optprobPowerAllocation_highSNR}
\end{align*}}\noindent
This can be solved and the solution is given by
\eqref{eq:highSNRsolution}.

In the low SNR $\bigl(\rho_{k^i}\lambda^{(k^i)}_{ij}\ll
\sigma_w^2\bigr)$, the cost function
\eqref{eq:optprobPowerAllocation} can be written as
{\small\begin{align*}
&\sum_{i:|\Kc_i|=1} \text{tr}\left(\frac{ a^{2(l M+M_p-k^i)} \sigma_w^2 \Lambdabf^{(k^i)}_i }{\rho_{k^i}\Lambdabf^{(k^i)}_i
+ \sigma_w^2\Ibf_{N_r}}  \right) \\
&= \sum_{i:|\Kc_i|=1} a^{2(l M+M_p-k^i)}\sigma_w^2
\text{tr}\left(\Ibf_{N_r} + \frac{ \Lambdabf^{(k^i)}_i - \sigma_w^2\Ibf_{N_r} }{\rho_{k^i}\Lambdabf^{(k^i)}_i + \sigma_w^2\Ibf_{N_r}} - \right.\\
&~~~ \left. \frac{ \rho_{k^i}\Lambdabf^{(k^i)}_i }{\rho_{k^i}\Lambdabf^{(k^i)}_i + \sigma_w^2\Ibf_{N_r}} \right) \\
&\simeq
\sum_{i:|\Kc_i|=1} a^{2(l M+M_p-k^i)} \text{tr}\left( \Lambdabf^{(k^i)}_i - \rho_{k^i}\Lambdabf^{(k^i)}_i \right), \\
&\Rightarrow~ \max_{\boldsymbol\rho} \sum_{i:|\Kc_i|=1} \rho_{k^i}
a^{2(l M+M_p-k^i)} \text{tr}\left( \Lambdabf^{(k^i)}_i \right).
\end{align*}}
This can be solved and the solution is given by
\eqref{eq:lowSNRsolution}.

\begin{spacing}{0.85}
\bibliographystyle{IEEEbib}
\bibliography{IEEEabrv,referenceBibs}
\end{spacing}

\end{document}